\documentstyle[aps,eqsecnum,psfig]{revtex}
\begin{document}
%\draft
% XXX: fixed use of macro
\newcommand{\epsfile}[1]{\psfig{#1}}
\title{ Calculation of the Flux of Atmospheric Neutrinos}

\author{M.~Honda, \hskip 5mm T.~Kajita,}
\address{Institute for Cosmic Ray Research, 
University of Tokyo, Tanashi, Tokyo 188, Japan}
\author{K.~Kasahara}
\address{Faculty of Engineering, Kanagawa University, Yokohama 221, Japan
}
\author{and}\address{}
\author{S.~Midorikawa,}
\address{
Faculty of Engineering, Aomori University, Aomori 030, Japan
}
\date{\today}
\maketitle
\begin{abstract}
Atmospheric neutrino--fluxes are calculated over the wide 
energy range from 30~MeV to 3,000~GeV
for the study of neutrino--physics using the data
from underground neutrino--detectors.
In this calculation, 
a full Monte Carlo method is employed for low energy neutrinos
(30~MeV -- 3~GeV),
while a hybrid method is used for high energy neutrinos (1 -- 3,000~GeV).
At low energies, the ratio $(\nu_e + \bar\nu_e)/(\nu_\mu + \bar\nu_\mu)$
agrees well with other calculations
and is significantly different from observations.
For the high energy neutrino--fluxes, 
the zenith angle dependence of atmospheric neutrino--flux is 
studied in detail, 
so that neutrino--oscillation parameters can be calculated %?
for comparison with experimental results.  %?
The atmospheric muon--flux at high altitude and 
at sea level is studied to calibrate the neutrino--fluxes at low energies
and high energies respectively.
The agreement of our calculation with observations is satisfactory.
The uncertainty of atmospheric neutrino--fluxes is also studied.
\end{abstract}
%\pacs{14.60.Pq, 95.85.Ry}
%\baselineskip 18pt
% XXX: save trees

\section{Introduction}
\label{sec:intro}

In this paper, we report the calculation of the atmospheric 
$\nu$--flux in the energy range from 30~MeV up to 3,000~GeV, 
corresponding to the observation range of underground neutrino
detectors. 
Detailed calculations of atmospheric $\nu$--fluxes are important, 
since the flux ratio, $\nu_\mu / \nu_e$, 
observed by many experiments 
shows a significant deviation from the expected 
value~\cite{hirata}\cite{casper}\cite{sudan2} 
at low energies ($<E_\nu> \sim 1$~GeV) 
and multi-GeV energies ($<E_\nu> \sim 5 - 7$~GeV)~\cite{fukuda}.
Many authors have considered the possibility that this deviation 
is evidence for $\nu$-oscillations, with a large mixing angle and 
$\Delta m^2 \sim 10^{-2}$~eV$^2$~\cite{hirata}\cite{learned}
\cite{berger}\cite{hhm}\cite{hhm2}.
The zenith angle variation of the
$(\mu/e)_{\rm data}/(\mu/e)_{\rm MC}$ ratio at multi--GeV energies 
is especially suggestive~\cite{fukuda}.
Above 10~GeV, up-going $\mu$'s are used to determine $\nu_\mu$--fluxes.
The variation of up-going $\mu$--fluxes with the arrival direction
can be used to study the oscillation parameters,
since the distance to the place where $\nu$'s are produced
is determined by the zenith angle of the arrival 
direction~\cite{bionta}\cite{oyama}\cite{frati}.

Atmospheric $\nu$--fluxes have been
calculated by Volkova~\cite{volkova}, Mitsui et al.~\cite{mitsui}, 
Butkevich et al.~\cite{butkevich} and Lipari~\cite{lipari} 
mainly for high energies 
(from around 1~GeV to above 100,000~GeV). 
Gaisser et al.~\cite{gaisser88}, 
Barr et al.~\cite{bar89}, 
Bugaev and Naumov~\cite{bn89}, Lee and Koh~\cite{lk90}, 
and Honda et al.~\cite{hkhm} calculated precisely the
atmospheric $\nu$--flux for low energies ($\lesssim 3$~GeV).
A calculation of low energy atmospheric $\nu$--flux
using the $\mu$--flux observed at high altitudes
has also been made~\cite{perkins}. 

In this paper, we use essentially the same models for particle 
interaction, atmospheric structure, and cosmic ray fluxes 
with Ref.~\cite{hkhm}. 
The calculation method, however, is different for low energy 
(30~MeV -- 3~GeV) and high energy (1 -- 3,000~GeV) atmospheric 
$\nu$'s.
We employ a full Monte Carlo method at low energies,
but use a hybrid method at high energies.

The difficulties 
in the calculation of atmospheric $\nu$--fluxes
differ between high and low energies.
In case of low energy $\nu$'s, the primary fluxes of cosmic ray 
components are relatively well known.
However, the low energy cosmic ray fluxes ($\lesssim$ 30~GeV)
are modulated by solar activity,
and are affected by the geomagnetic field through
the rigidity ($=$ momentum$/$charge) cutoff.
For high energy $\nu$'s ($>$ 100~GeV), 
the $\gtrsim$ 1,000~GeV cosmic ray flux is relevant.
At these energies, solar activity and rigidity cutoff 
do not affect cosmic rays,
but details of the cosmic ray flux are not as well--measured.
Details of the hadronic interactions of cosmic rays with 
air nuclei are also a source of the uncertainty
in the calculated $\nu$--fluxes.
At low energies, the proton--nucleus interaction
at $\lesssim$ 30~GeV is important.
There have been many accelerator experiments studying hadronic 
interactions in this energy region, however, 
not many are suitable for our purpose.
In the high energy proton-proton interactions, 
it is normally assumed that
the spectrum of secondary particles satisfies 
the Feynman scaling hypothesis,
which is confirmed by collider experiments up to a lab.
energy of 3,000~TeV.
Although there is a weak breaking of the hypothesis  
in the central region,
it has no significant effect on atmospheric 
$\nu$--fluxes.

We employ an one--dimensional approximation in which
all the secondary particles and the $\nu$'s keep 
the direction of their parent cosmic rays, 
throughout the energy range of concern.
For high energy $\nu$'s, this is a good approximation
because of the nature of hadronic interactions.
We also expect that the directional--average of 
$\nu$--fluxes may be calculated with good accuracy 
even at low energies.
However, when we need information about the variation with
direction of the low energy atmospheric $\nu$--flux,
especially for near horizontal directions, 
a three--dimensional calculation is necessary.
At low energies, secondary particles are produced 
with large scattering angles by hadronic interactions,
and the curvature of low energy $\mu$'s due to the geomagnetic 
field becomes sizable.
We note that a three--dimensional calculation of $\nu$--fluxes
with the Monte Carlo method requires an enormous computation 
time, 
since we need to calculate the $\nu$--flux at every position 
on the Earth, and for all directions with good statistics.

Section~\ref{sec:primary} is devoted to the problems with 
low-energy primary cosmic rays (\ref{sec:1ry-low}), 
such as the effect of solar activity (\ref{sec:modulation})
and the rigidity cutoff due to the geomagnetic field 
(\ref{sec:cutoff}).
Also in~\ref{sec:1ry-high}, 
the primary cosmic ray fluxes are compiled for 
each chemical composition
in the energy region of 100~GeV -- 100~TeV,
for use in the calculation of atmospheric $\nu$'s.
The processes which take place during the propagation of cosmic rays 
in the atmosphere are explained in section~\ref{sec:int-in-air}.
The hadronic interaction model we employ is 
explained in~\ref{sec:hadronic-int}.
The decay of mesons, such as
$\pi$'s and $K$'s, which are created in cosmic ray interactions,
is the main source of the atmospheric $\nu$'s.
These decay processes 
are summarized in section~\ref{sec:decay} with a discussion
of muon polarizations.
In section~\ref{sec:nflx}, we explain the 
calculation of atmospheric $\nu$--fluxes,
and the results are summarized in~\ref{sec:nflx-low} for 
30~MeV -- 3~GeV,
and in~\ref{sec:nflx-high} for 1~GeV -- 3,000~GeV.
In section~\ref{sec:mflx}, 
atmospheric $\mu$--fluxes are calculated by the same 
method as the $\nu$--fluxes, and are compared with the
observed data.
In section~\ref{sec:syserror}, 
the uncertainties in the calculation of
atmospheric $\nu$--fluxes are discussed.
In section~\ref{sec:summary}, 
the major results of this work are summarized. 

\begin{figure}
\centerline{\epsfile{file=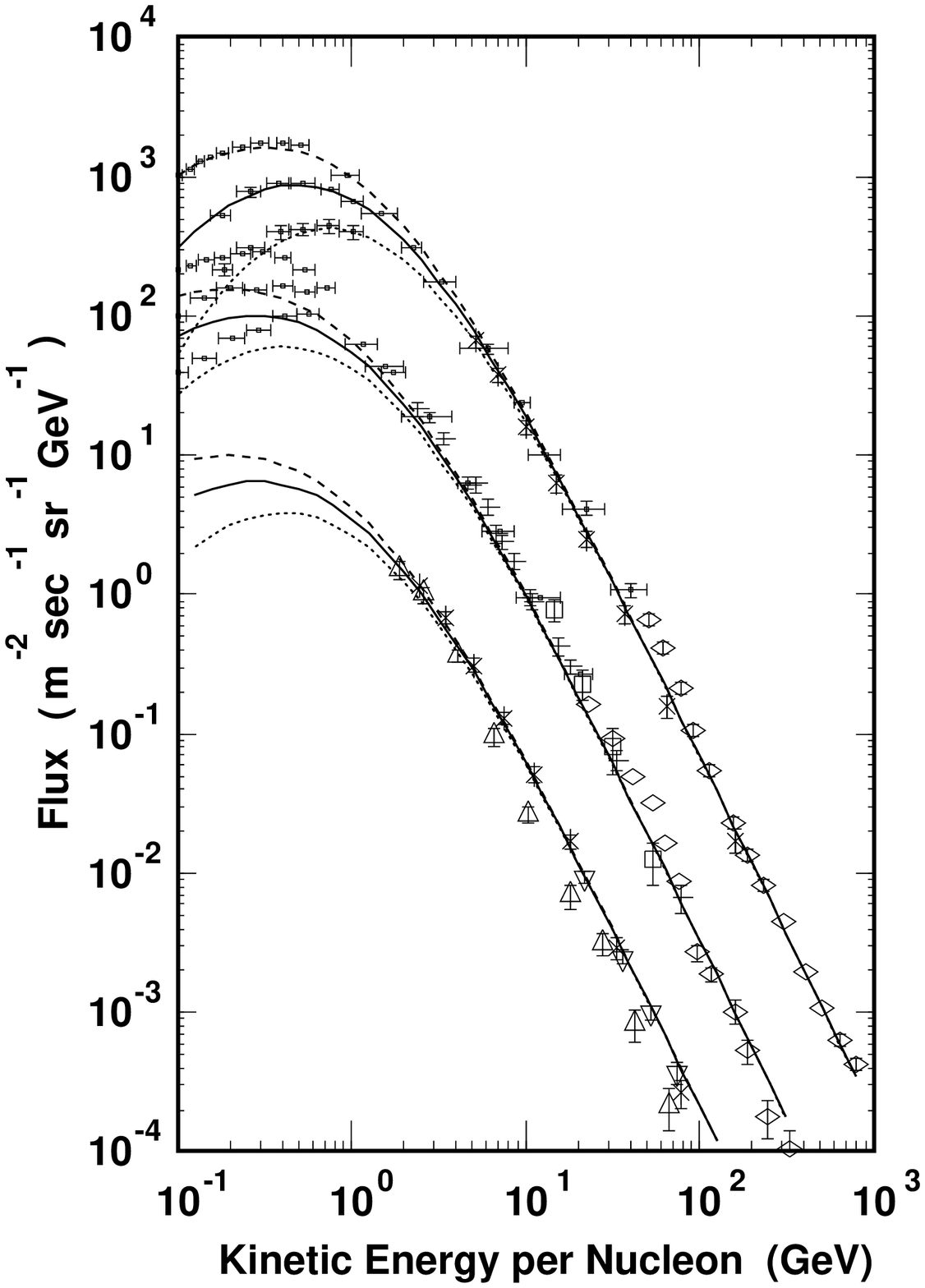,height=10cm}}
\caption{}
\vspace{5mm}
{\baselineskip = 12pt
Observed fluxes of cosmic ray protons, helium nuclei, and CNOs from
the compilation of Webber and Lezniak~[22]. %[wl1]
(Dots represent data from Refs~[23] %[ormes-webber]
and [24], %[wlkd]
diamonds from 
Ref.~[25], %[ryan]
crosses from Ref.~[26], %[smith]
minuses from Ref.~[27], %[ananda]
squares from Ref.~[28], %[verma]
upward triangles from Ref.~[29], %[bala]
and downward triangles from Ref.~[30]. %[juli]
)
Solid lines are our parametrization for solar mid.,
dash lines for solar min., and dotted lines for solar max.
}
\label{fig:1ry-low}
\end{figure}

\section{The Flux of Cosmic Rays}
\label{sec:primary}
\subsection{The Flux of Cosmic Rays below 100 GeV}
\label{sec:1ry-low}

Primary cosmic ray fluxes are relatively well known in the
low energy region ($\lesssim 100$~GeV), by which the low energy
atmospheric $\nu$--fluxes ($\lesssim$ 3~GeV) are mainly created.
However, the fluxes are affected by solar
activity and the geomagnetic field.  
The effect of solar activity is known as the solar-modulation 
of cosmic rays, and is commonly parametrized by the sun-spot 
number or the count rate of neutron monitors.  
The effect of the geomagnetic field is represented as the 
rigidity cutoff of cosmic rays. 
In the following, the treatment of these effects in this 
calculation is explained.

Webber and Lezniak have compiled the energy spectrum of primary cosmic
rays for hydrogen, helium, and CNO nuclei in the energy range 
10~MeV $\sim$ 1,000~GeV \cite{wl1} for three levels of solar 
activity (Fig.~\ref{fig:1ry-low}).  
A similar compilation has been made by Seo et al.~\cite{seo1} 
for hydrogen and helium nuclei, which agrees well with that 
of Webber and Lezniak.  
Seo et al. estimated that uncertainties in the instrumental 
efficiency ($\sim$ 12\%) and exposure factor  (2 -- 3\%)
result in the overall uncertainty of the primary cosmic ray 
fluxes being $\sim$ 15\%~\cite{seo1}.

From the compilation of Webber and Lezniak, 
the chemical composition of cosmic rays is H(proton) 
$\sim$ 90.6~\%, He$\sim$ 9.0~\%, and CNO nuclei $\sim$ 0.4~\% 
above $\sim 100 $~MeV$/$nucleus, 
and H $\sim$ 95.2~\%, He$\sim$ 4.5~\%, and CNO nuclei 
$\sim$ 0.3~\% above $\sim$ 2~GeV$/$nucleus.  
The portion of other components (Ne, S, Fe, $\cdots$) is
so small that they can be neglected in the calculation of 
low-energy atmospheric $\nu$--fluxes.  
It is noted that atmospheric $\nu$'s are created
through the hadronic interactions of cosmic rays and air nuclei,
and therefore are dependent on the number of nucleons rather 
than the number of nuclei.  
The contribution of a heavier cosmic ray nucleus to the 
atmospheric $\nu$--flux is larger than that of a cosmic ray proton.
Helium nuclei carry $\sim$ 15~\% of the total nucleons in the 
cosmic ray flux and the
CNO group carries $\sim$ 3.6~\% above $\sim$ 2~GeV$/$nucleon.  
These effects
are amplified by the effect of the geomagnetic field through
the rigidity cutoff, 
since the rigidity for those nuclei is two times larger
than protons with the same momentum$/$nucleon.  
The details are given below.

\subsection{Solar Modulation}
\label{sec:modulation}

The flux of low energy cosmic rays is modulated by solar activity.
The solar wind drives back the low energy cosmic
rays which are entering into the solar sphere of influence,
and the strength of the solar wind varies with solar activity.  
This effect is more evident in the lower energy cosmic rays:
the flux difference at solar maximum and solar minimum is 
more than a factor of two for 1~GeV cosmic rays,
and it decreases to $\sim$ 10~\% for 10~GeV cosmic rays 
(Fig.~\ref{fig:1ry-low}).

The primary flux for various levels of solar activity is parametrized 
by Nagashima et al.~\cite{nagashima} as a function of the count rate
of a neutron monitor $N$ by
\begin{equation}
f(E_k)dE_k = \gamma_i u^{-2.585} M(p, N) dE_k,
\label{eq:1ryeq1}
\end{equation}
where $i$ stands for the kind of nucleus ($=$H, He, CNO ..), 
$p$ for rigidity in GV ($\equiv$GeV/c/Z), 
$E_k$ for kinetic energy per nucleon in GeV, 
and $u$ for total energy per nucleon in GeV.  
The absolute flux value for each component is determined by $\gamma_i$,
where
$\gamma_{\rm H}=10.85\times 10^3$ ${\rm m^2 sec^{-1} sr^{-1}
GeV^{-1}}$, 
$\gamma_{\rm He}=
5.165\times 10^2$ ${\rm m^2 sec^{-1} sr^{-1} GeV^{-1}}$, 
and 
$\gamma_{\rm CNO}=3.3 \times 10^{-2}$ ${\rm m^2 sec^{-1} sr^{-1}
GeV^{-1}}$ respectively.  
The function $M(p,N)$ is the modulation function defined by
\begin{equation}
M(p, N) = \exp[ - {1.15+14.9 (1-N/N_{max})^{1.12} \over 0.97 +
(p/1GV)}],
\label{sec:1ryeq2}
\end{equation}
and $N$ is the count rate of the neutron monitor at Mt.~Washington 
with $N_{max}=2465\ {\rm count/hour}$.  
We take $N=2445$ for solar min., $N=2300$ for solar mid., 
and $N=2115$ for solar max.
The results of the parametrization are shown in 
Fig.~\ref{fig:1ry-low} by the solid, dashed and dotted lines, 
which agree well with data except for low energy helium nuclei
$(\lesssim 10~{\rm GeV/nucleon})$. 
However, 
this produces only a very small effect on the calculation of 
atmospheric $\nu$'s due to the proton dominance of the 
cosmic ray flux and the relatively small contribution of 
cosmic rays of this energy.

\subsection{Rigidity Cutoff of the Geomagnetic field}
\label{sec:cutoff}

The geomagnetic field determines the minimum energy with which a
cosmic ray can arrive at the Earth.  
This effect is caused by the magnetic shield effect for low 
energy cosmic rays. 
For the cosmic ray nucleus, the minimum energy of cosmic rays
arriving at the Earth is determined by the minimum rigidity 
(rigidity cutoff) rather than the minimum momentum.  
We note that the rigidity cutoff is a function of the entering 
position on the Earth and arrival direction 
(zenith angle, $\theta$, and azimuth angle, $\phi$).  
Since the mass$/$charge ratio of helium and CNO nuclei is twice 
that of protons, those nuclei carry lower energy
nucleons into the atmosphere than protons.

The actual geomagnetic field is represented by a multipole expansion
of the spherical harmonic function as
\begin{equation}
B_{North} = {1 \over r} {\partial V \over \partial \theta}, \ \ 
B_{East} = -{1 \over \sin\theta} {\partial V \over \partial \phi},
\ \ {\rm and}\ \ 
B_{Down} = {\partial V \over \partial r}
\label{eq:magfield}
\end{equation}
with the potential function
\begin{equation}
V = R \sum_{n=1} \sum_{m=0}^{n} ({R \over r})^{n+1}
[g_n^m \cos(m\phi) + h_n^m \sin(m\phi)]
P_n^m(\cos\theta).
\label{eq:potential}
\end{equation}
Here $R$ is the radius of the Earth and
$P_n^m(x)$ is the associated Legendre function.
The expansion coefficient $(g_n^m, h_n^m)$ is compiled and
reported by IAGA Division I working Group 1~\cite{gmtable}.
As the geomagnetic field varies slowly with time,
the coefficient is reported yearly with the time differential
values.

The value of the rigidity cutoff for the actual geomagnetic 
field can be obtained from a computer simulation of 
cosmic ray trajectories.
In this simulation, an antiproton,
which has the same mass as a proton but
the opposite charge, 
is used as the test particle.
We note that the change $e \leftrightarrow -e$ is equivalent to
the change of $t \leftrightarrow -t$ in the equation of motion
of a charged particle in a magnetic field;
\begin{equation}
{\partial {\bf p} \over \partial t} = e {\bf v} \times {\bf B}
\label{eq:eq-motion},
\end{equation}
where {\bf p} is the momentum, 
{\bf v} is the velocity,
and {\bf B} is the magnetic field.
To determine the rigidity cutoff at different positions 
and for different directions,
we launch antiprotons from the earth,
varying the position and direction.
When a test particle with a given momentum
reaches a distance of 
10 times the Earth's radius, where
the strength of the geomagnetic field decreases to 
the same level as the interstellar magnetic field 
($\sim 3\times 10^{-8}$~T),
it is assumed that the test particle has escaped from 
the geomagnetic field.
Assuming the momentum distribution of 
cosmic rays is isotropic in angular space,
some cosmic rays which have the 
same rigidity can arrive at the Earth
following the same trajectory but in the opposite
direction.
The rigidity cutoff is calculated as the
minimum momentum with which the test particle escapes from the
geomagnetic field.
We note that for protons the rigidity and the momentum is the
same quantity.

\begin{figure}
\centerline{\epsfile{file=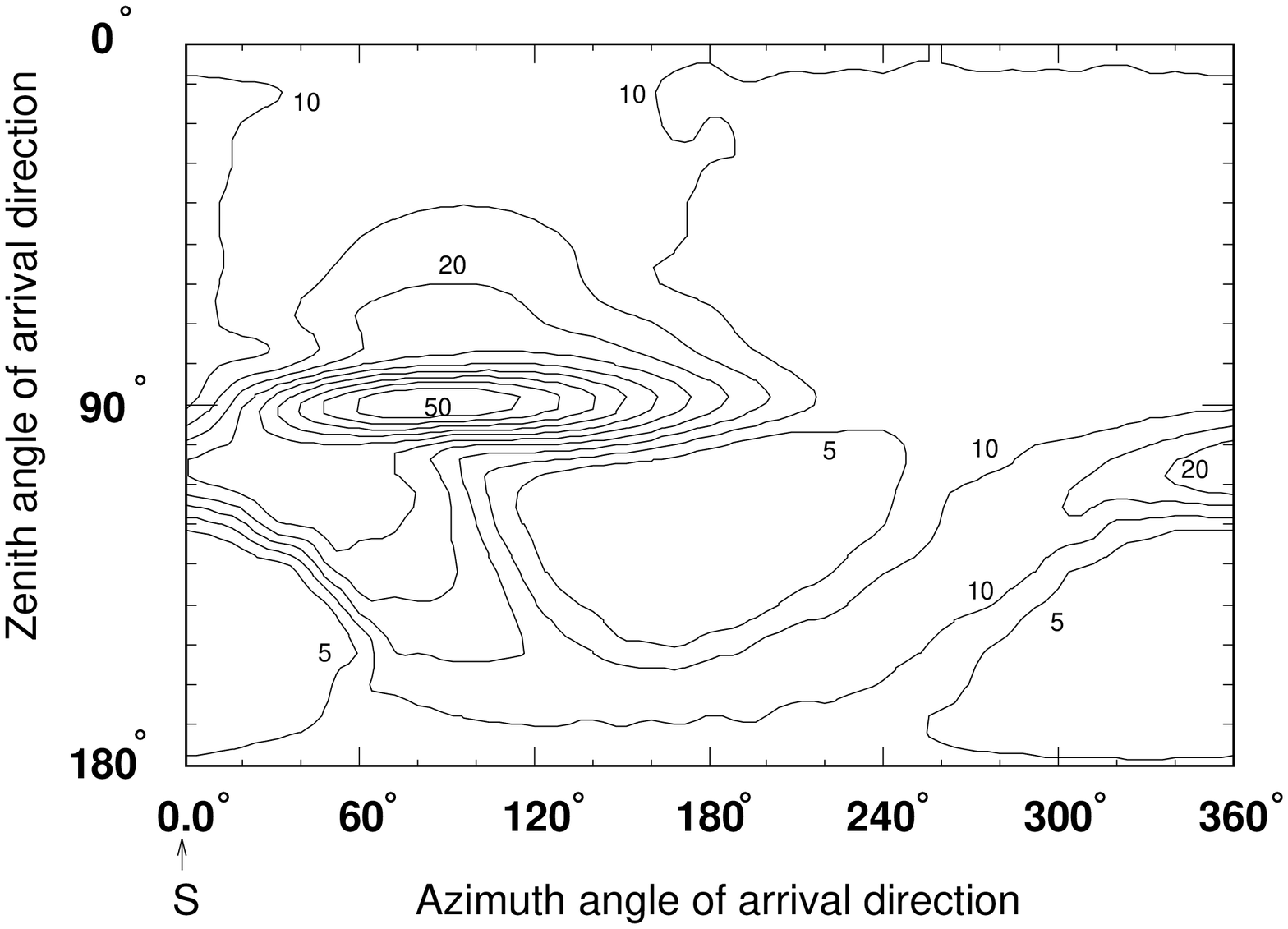,height=6.5cm}}
\caption{}
\vspace{5mm}
{\baselineskip = 12pt
The contour map of cutoff-rigidity for the $\nu$ arrival 
directions at Kamioka.
Azimuth angles of $0^\circ, 90^\circ, 180^\circ,$ and $270^\circ$
show directions of 
south, east, north, and west respectively.
}\label{kamcut}
\end{figure}

In an one--dimensional approximation,
we need the  rigidity for the arrival direction of $\nu$'s
at $\nu$--detector sites.
We found that the magnetic field calculated up to 5th order
of the expansion
gives almost the same result for rigidity cutoff 
as calculations with higher order expansions.
The rigidity cutoff at Kamioka is 
shown as a contour map in Fig.~\ref{kamcut}. 
We note that the dipole approximation is good near the
magnetic equator.
However, the multipole effect becomes important 
near the magnetic pole.
For the calculation of atmospheric $\nu$--fluxes, we need the 
multipole expression for the geomagnetic field even when 
the detectors are not located near the pole,
since $\nu$'s created near the pole arrive at the
detector through the Earth.

\subsection{Cosmic Ray Fluxes above 100~GeV}
\label{sec:1ry-high}

Cosmic rays with energy greater than 100~GeV, 
which are responsible for $\gtrsim$ 10~GeV atmospheric 
$\nu$--fluxes, are not affected by
solar activity or by geomagnetic effects.
However, ther are few measurements of the cosmic ray chemical 
composition at these energies, especially above 1~TeV.
Here, we compile the available data of cosmic ray fluxes 
for H, He, CNO, Ne-S, and Fe-group nuclei up to 100~TeV$/$nucleon.
Above 100~TeV$/$nucleon, the cosmic ray spectrum is measured by
the air shower technique and almost no direct measurements of 
cosmic ray particles are available.

\begin{figure}
\centerline{\epsfile{file=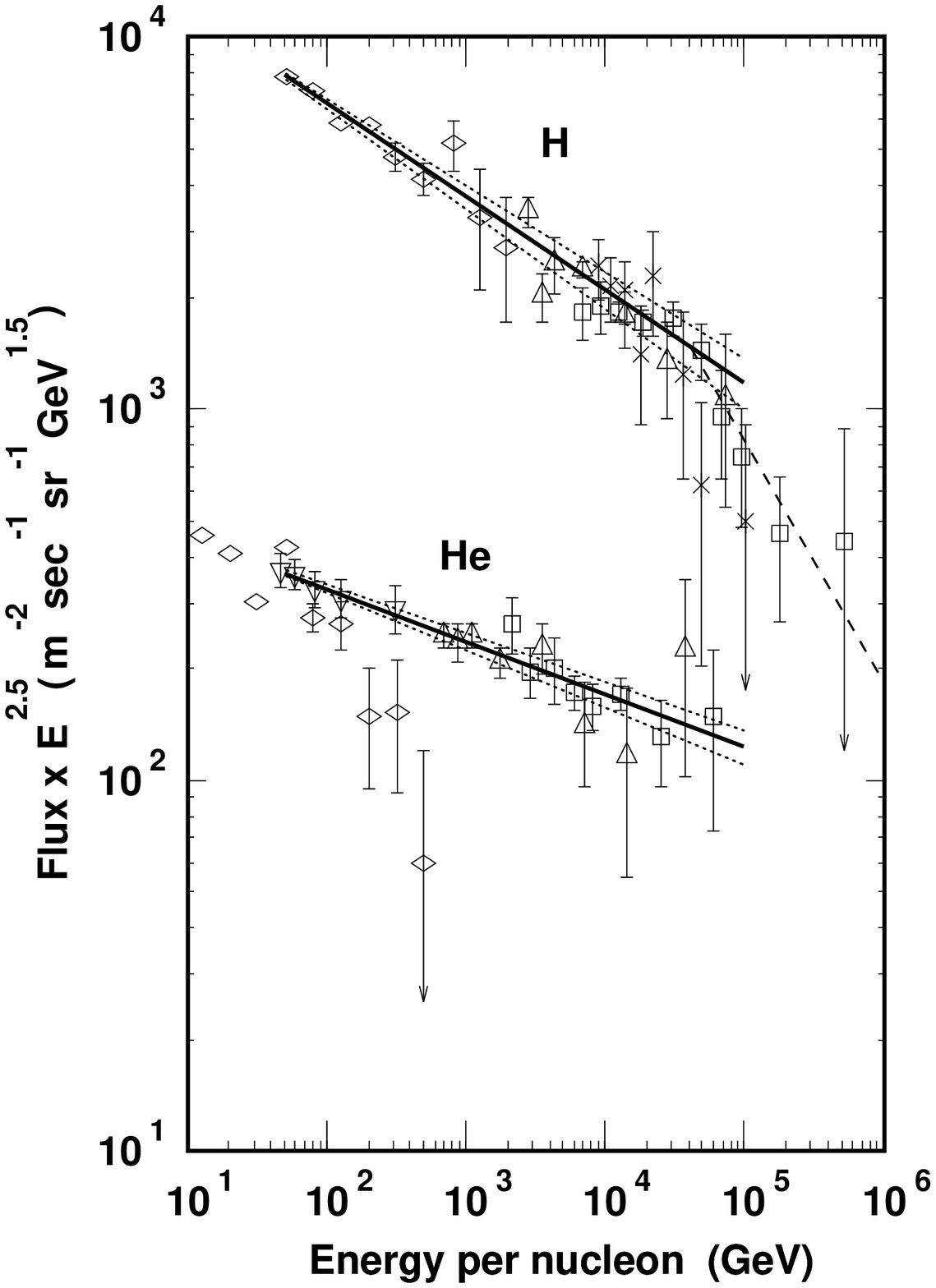,height=10cm}}
\caption{}
\vspace{5mm}
{\baselineskip = 12pt
The observed cosmic ray fluxes for H and He nuclei.
The lines show our fits, as explained in the text.
Diamonds are from Ref.~[25], %[ryan]
squares from Refs~[34] and [35], %[jacee-p] and [jacee-nucl]
upward triangles from Ref.~[36], %[ivanenko]
downward triangles for He from Ref.~[37], %[chicago-he]
and crosses for H from Ref.~[38]. %[aoyama]
The dashed line show the JACEE spectrum for H
above 40~TeV~[35].
}
\label{fig:p-he}
\end{figure}

In Fig.~\ref{fig:p-he},
observed cosmic ray fluxes from 
Refs~\cite{ryan}\cite{jacee-p}\cite{ivanenko}
\cite{chicago-he}\cite{aoyama}
are summarized for H and He.
We fitted the observed flux for $\ge$ 100~GeV with
a single power function:
\begin{equation}
Flux(E) = A \cdot (E/100~{\rm GeV})^\gamma \ ,
\end{equation}
and show the result in the same figure.
We note that the data for He of Ryan et al.~\cite{ryan} are more 
than two times smaller than those of other groups 
and their error bars are larger than others.
Therefore, we have not used their data in this analysis.
The observed cosmic ray flux for CNO, Ne-S, and Fe-group nuclei 
from 
Refs~\cite{smith}\cite{jacee-nucl}\cite{aoyama}
\cite{simon}\cite{juli}\cite{heao-3}\cite{crn}
are 
shown in Fig.~\ref{fig:c-fe}, also with our fitted spectra.
The parameters ${A, \gamma}$ are summarized in 
Table~\ref{table:fit-parm} for H, He, CNO, Ne-S, and Fe-group 
nuclei.

\begin{figure}
\centerline{\epsfile{file=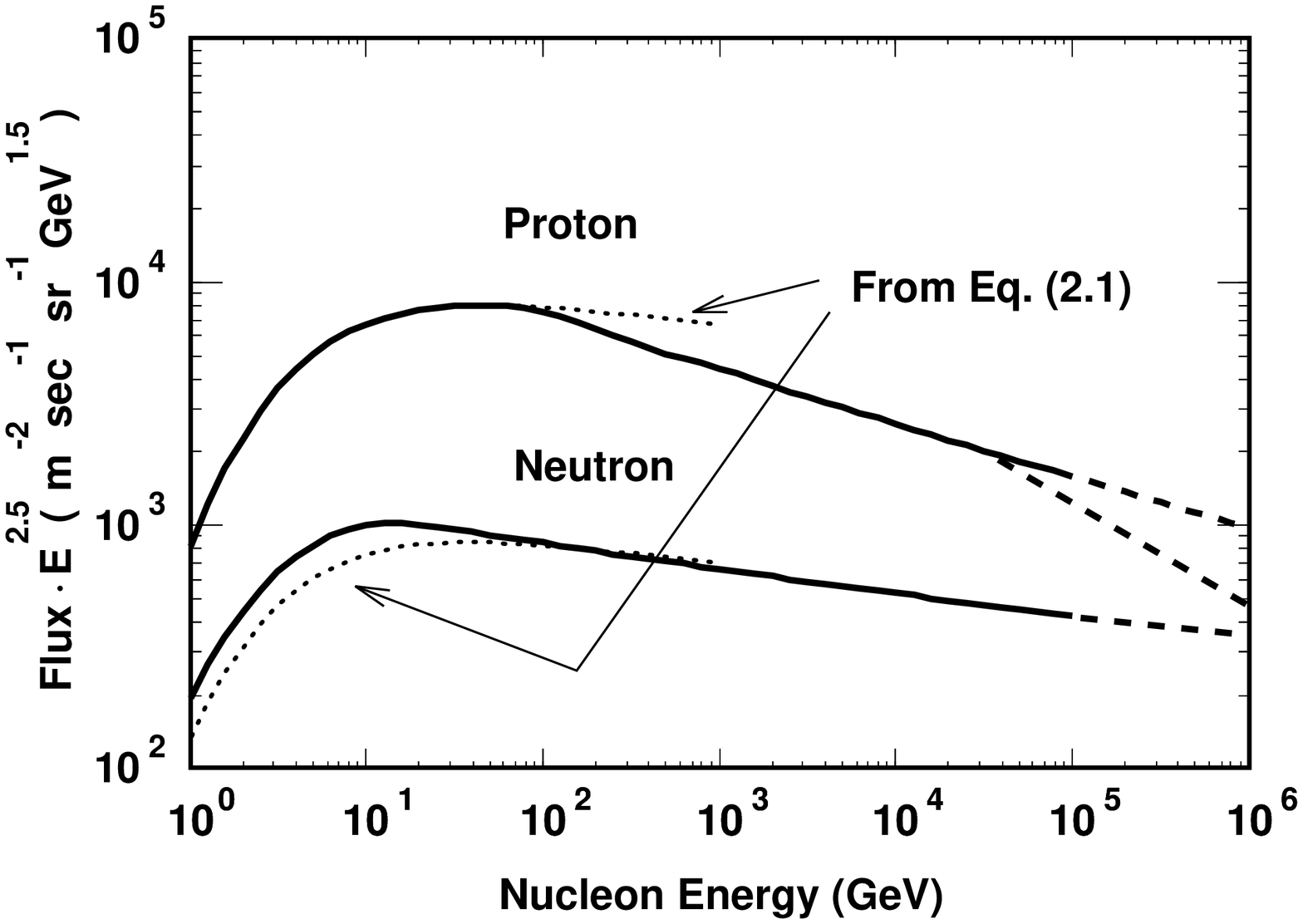,height=6.5cm}}
\caption{}
\vspace{5mm}
{\baselineskip = 12pt
The cosmic ray nucleon flux summed over all chemical components,
and extrapolated (dashed line) up to 1,000~TeV.
The lower line for protons above 40~TeV assumes
the steepening suggested by the JACEE group.
}
\label{fig:nucleon}
\end{figure}

\begin{figure}
\centerline{\epsfile{file=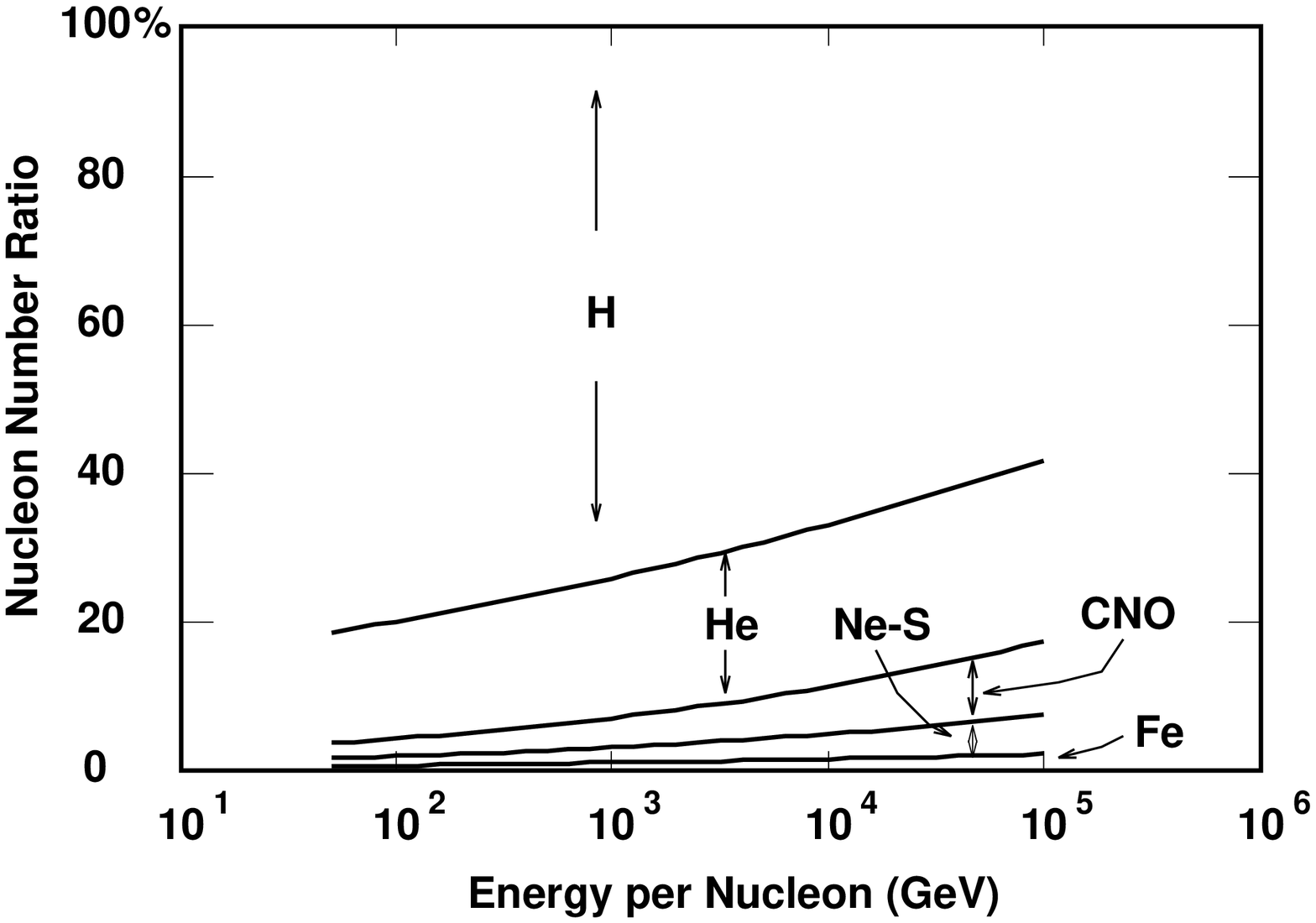,height=6.5cm}}
\caption{}
\vspace{5mm}
{\baselineskip = 12pt
Ratios of nucleons carried by H, He, CNO, Ne-S, and Fe-group
nuclei.
}
\label{fig:1ry-ratio}
\end{figure}

For the calculation of the atmospheric $\nu$--fluxes
of energy region 1 -- 3,000~GeV, 
we employed the superposition approximation.
This approximation treats a nucleus as a bundle of 
independent nucleons,
and considers the event caused by the nucleus as the sum of 
independent events caused by these nucleons.
Therefore,
we need the flux of each nucleon rather than that of 
each chemical composition. 
In Fig.~\ref{fig:nucleon}, 
we depicted the nucleon flux (proton and neutron) 
calculated from the single power function fit for 
H, He, CNO, Ne-S, and Fe-group cosmic rays.
For $< 100$~GeV, we grouped the experimental points 
in Figs~\ref{fig:1ry-low},~\ref{fig:p-he}, and~\ref{fig:c-fe} 
in several energy bins and calculated the average flux over 
different groups.
Thus calculated proton flux below 100~GeV agrees well 
with the parametrization of equation~\ref{eq:1ryeq1} 
with $N=2300$ (solar mid.), while the neutron flux is larger.
This difference has already appeared in Fig.~\ref{fig:1ry-low}, 
but produces very small effects on 
the atmospheric $\nu$--flux as explained before.
It is noted that the proton cosmic rays constitute $\sim$ 80~\% 
of all nucleons at 100~GeV$/$nucleon, and helium $\sim 15$~\%.
This percentage decreases with energy,
however, 
$\sim$ 80~\% of nucleons are carried by proton and helium cosmic rays
even at 100~TeV$/$nucleon (Fig.~\ref{fig:1ry-ratio}).
The heavier nuclei are still a minor component of cosmic rays 
at this energy.

\begin{figure}
\caption{}
\vspace{5mm}
{\baselineskip = 12pt
Observed cosmic ray fluxes for CNO, Ne-S, and Fe nuclei.
The lines show the fitted result as is explained in the text.
Pluses are from Ref.~[26], %[smith]
minuses from Ref.~[30], %[juli]
squares from Ref.~[35], %[jacee-nucl]
crosses from Ref.~[38], %[aoyama]
dots from Ref.~[39], %[simon]
upward triangles from Ref.~[40], %[heao-3]
downward triangles from Ref.~[41], %[crn]
and
diamonds from  Ref.~[42]. %[chicago-cno]
}
\label{fig:c-fe}
\end{figure}

\begin{figure}
\centerline{\epsfile{file=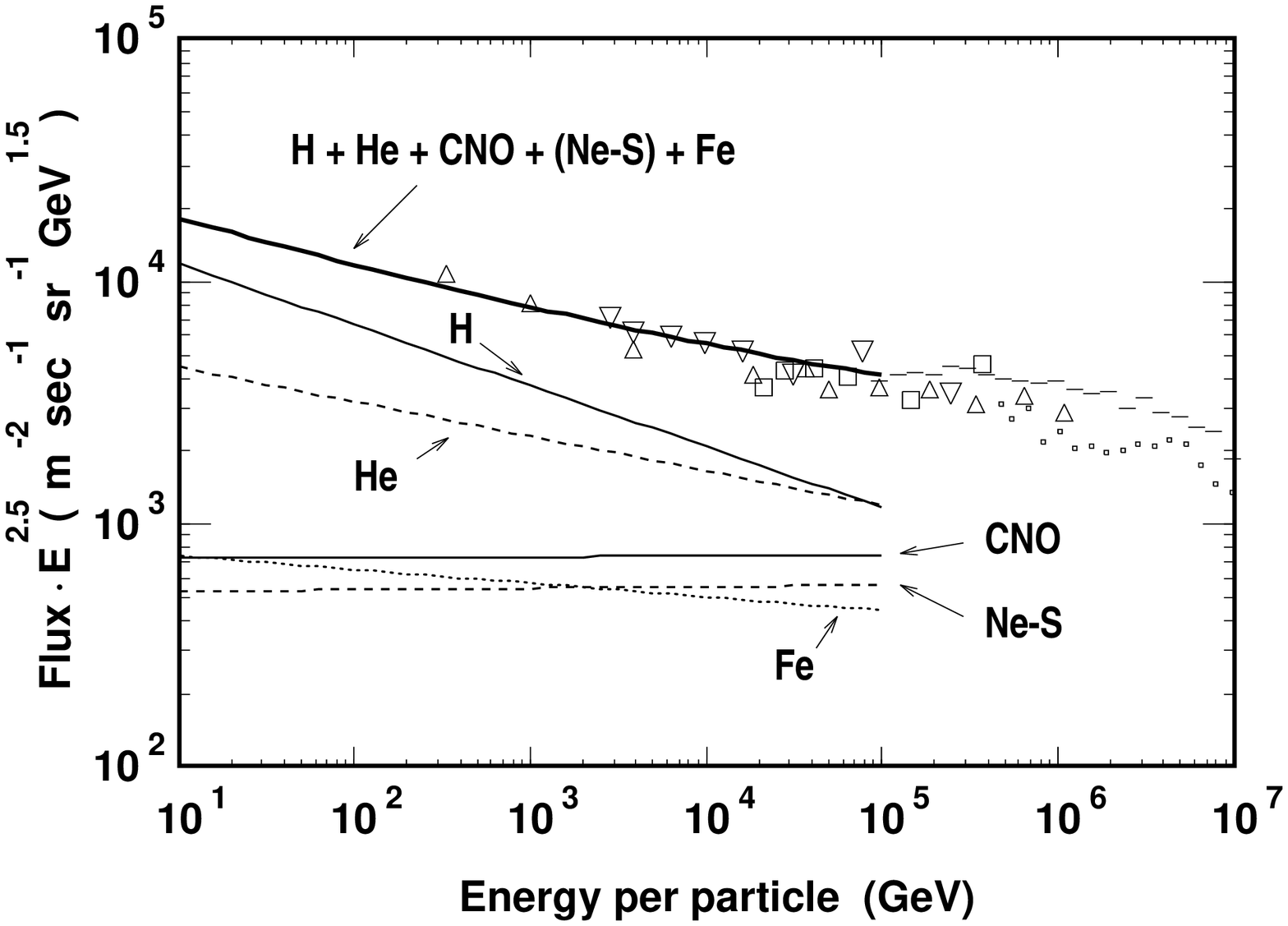,height=6.5cm}}
\caption{}
\vspace{5mm}
{\baselineskip = 12ptThe all-particle flux summed over 
fitted spectra for H, He, CNO, Ne-S, and Fe-group nuclei
compared with the observed data.
Large squares from Ref.~[35], %[jacee-nucl]
downward-triangles from Ref.~[36], %[ivanenko]
small squares are from Ref.~[43], %[akeno]
minuses from Ref.~[44], and %[chacaltaya]
upward-triangles from Ref.~[45]. %[proton]
}
\label{fig:particle}
\end{figure}

Above 100~TeV, 
almost no observations of the cosmic ray
chemical composition are available.
However, the all-particle flux is well measured by 
the air-shower technique (e.g. see Ref.~\cite{akeno}).
Converting the energy-per-nucleon spectra back to 
energy-per-particle,
the H, He, CNO, Ne-S, and Fe-group nuclei fluxes were summed
and compared with the observed all-particle spectrum~\cite{jacee-nucl}
\cite{ivanenko}\cite{akeno}\cite{chacaltaya}\cite{proton}
in Fig.~\ref{fig:particle}. 
It is seen that the extrapolation of the calculated all-particle 
flux agrees well with the experimental data even at energies 
above 100~TeV.
We used the nucleon flux calculated here up to 1,000~TeV
(upper extrapolation for protons in Fig~\ref{fig:nucleon}).
The JACEE group suggested that the spectrum of cosmic protons
becomes steeper ($\gamma=3.22$) above 40~TeV~\cite{jacee-p}
(See Fig.~\ref{fig:p-he}).
The effect of this steepening has been studied 
for $\nu$--fluxes at 1,000~GeV, but is very small.
Errors in the cosmic ray nucleon spectra obtained here are
estimated to be $\lesssim$ 10~\% at around 100~GeV and increase to
$\sim$ 20~\% at 100~TeV.

\section{The Interaction of Cosmic Rays in Air}
\label{sec:int-in-air}
\subsection{Hadronic Interaction}
\label{sec:hadronic-int}

As cosmic rays propagate in the atmosphere,
they create $\pi$'s and $K$'s in interactions with air nuclei.
These mesons create atmospheric $\nu$'s when they decay
as follows:
\begin{equation}
\begin{array}{lllll}
A_{cr} + A_{air} \rightarrow  &\pi^\pm, K^\pm, K^0, \cdots\\
&\pi^+   \rightarrow &\mu^+ + \nu_\mu& & \\
&&\mu^+  \rightarrow &e^+ + \nu_e + \bar \nu_\mu\\
&\pi^-   \rightarrow &\mu^- + \bar \nu_\mu& & \\
&&\mu^-  \rightarrow &e^- + \bar \nu_e + \nu_\mu\\
&.\\
&.\\
\end{array}
\end{equation}
The interactions of cosmic ray protons and nuclei with 
air nuclei are hadronic in character.
We employed the NUCRIN~\cite{nuc1} Monte Carlo code
for hadronic interactions for lab. energies $\le$ 5~GeV,
and LUND code -- FRITIOF version 1.6~\cite{lund1}
and JETSET version 6.3~\cite{lund2} -- for 
$ 5~{\rm GeV} \le E_{lab} \le 500~{\rm GeV}$.
Above 500~GeV, 
the original code developed by Kasahara et al.
(COSMOS)~\cite{kasahara} was used.
The $K/\pi$ ratio
is taken 7~\% at 10~GeV, 11~\% at 100~GeV, and 14~\%
at 1,000~GeV in lab. energy.
We compared the output of those codes 
with available experimental data.
Although there have been many accelerator experiments of 
proton--nucleus and Helium-nucleus interactions 
for $\lesssim$ 30~GeV, 
not many are applicable to our present purpose.
For higher energies, the data of collider experiments are 
available for $p-p$ and $p-\bar p$ collisions.
We estimate the error of the atmospheric $\nu$ calculation
resulting from these Monte Carlo codes is around 10~\%.

\begin{figure}
\centerline{\epsfile{file=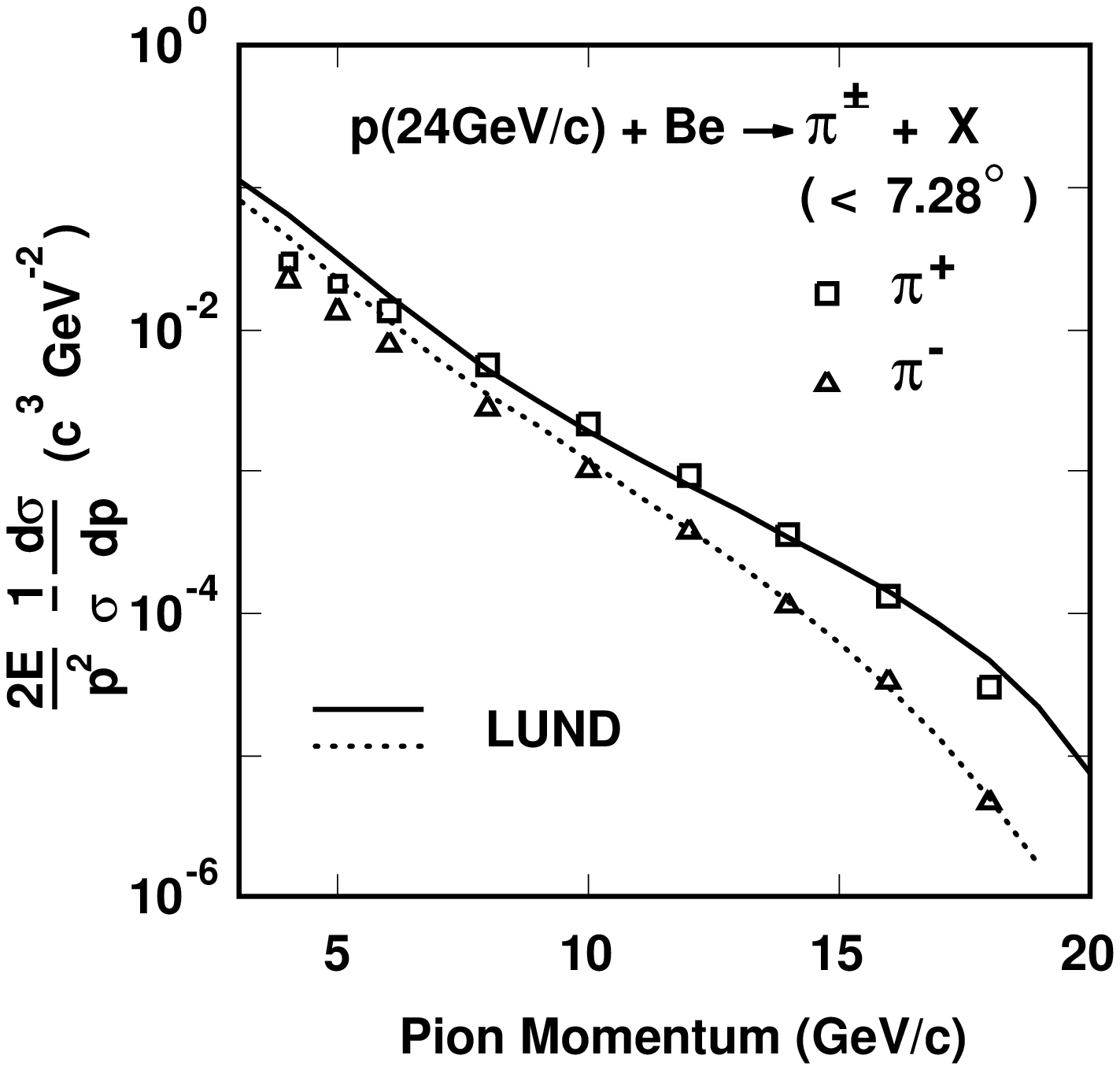,height=6.5cm}}\
\caption{}
\vspace{5mm}
{\baselineskip=12pt
Comparison of
$\pi^\pm$--production spectrum in $p-Be$
collisions at $p=24$ GeV$/$c
between experiment (Ref.~[50]) %[p24be] 
and the LUND Monte Carlo code.
The direction of the $\pi$'s are limited to $\leq 7.28^\circ$ 
from the direction of projectile protons.
}\label{p24be}
\end{figure}

\begin{figure}
\centerline{\epsfile{file=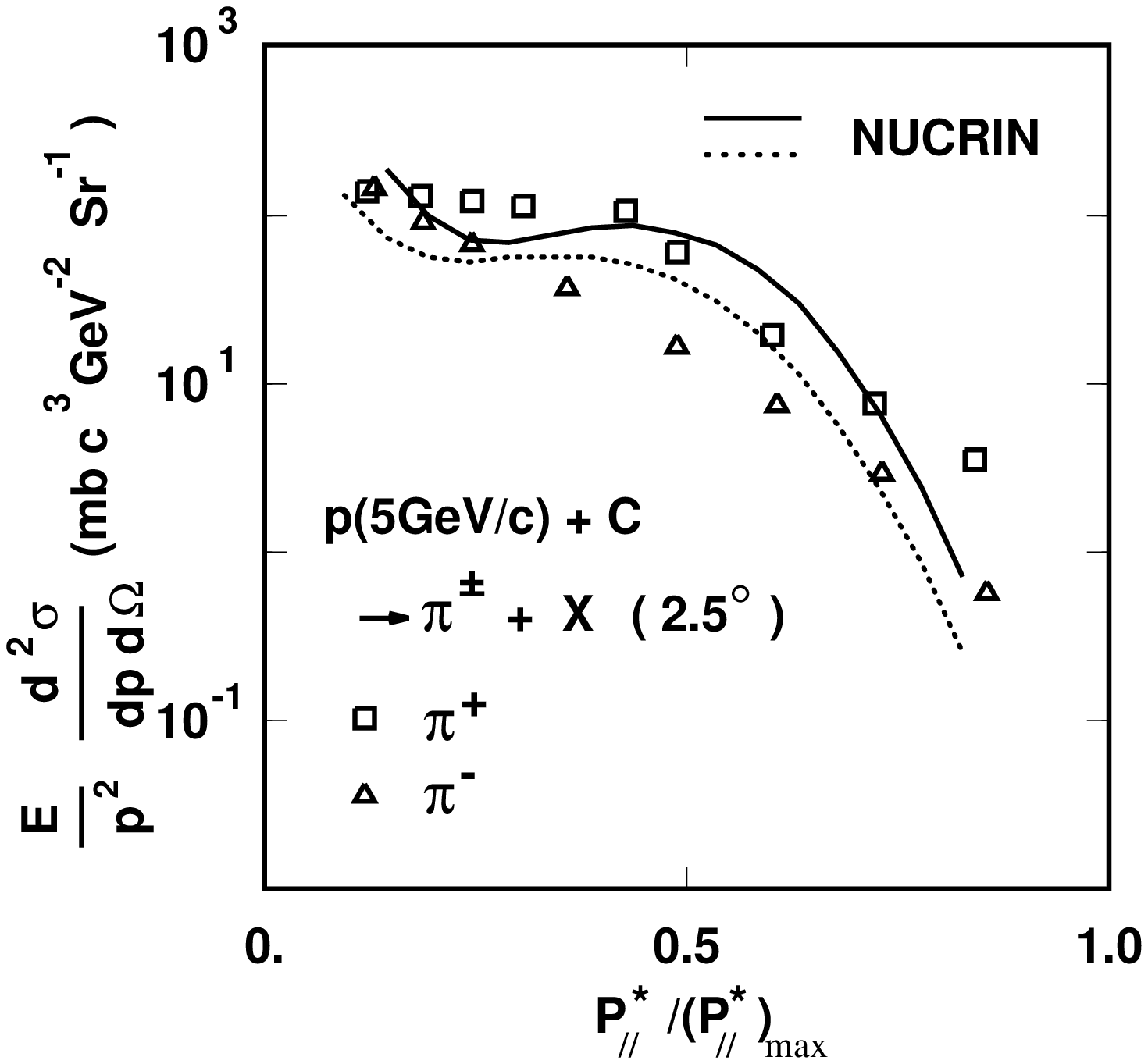,height=6.5cm}}
\caption{}
\vspace{5mm}
{\baselineskip=12pt
Comparison of
$\pi^\pm$--production spectrum in $p-C$ collisions at 
$p=5.1$GeV to the direction of $2.5^\circ$ between 
experiment (Ref.~[51])%[p5c]
and the NUCRIN Monte Carlo code.
$P^*$ denotes the momentum in the total center of 
mass system.
}\label{p5c}
\end{figure}

Results from the LUND code were compared with the
$\pi$ production spectrum in a cone of $\leq 7.28^\circ$
in $p-Be$ collisions (24 GeV$/$c)~\cite{p24be} (Fig.~\ref{p24be}).
NUCRIN results were compared with the $\pi$ production 
spectrum at 2.5$^\circ$ in $p-C$ collisions
(5 GeV$/$c)~\cite{p5c} (Fig.~\ref{p5c}).
The agreement of the LUND code and experimental data is 
quite good ($\sim$ 10~\%) except for the very low momentum 
region ($\lesssim 5$ GeV).
Since the energy spectrum of cosmic rays is steep,
the spectrum of $\pi$'s production by nucleons in the 
lower momentum region is relatively unimportant.
The agreement of the NUCRIN code and experimental data
seems not as good as that of the LUND code.
The authors of the NUCRIN code estimated the disagreement 
of their output and experimental result
as 10 -- 20~\%~\cite{nuc2}.
In low energy cosmic ray interactions,
the detailed structure of the $\pi$ production spectrum
may be smeared due to the flattening of the cosmic ray spectrum
at low energies.

\begin{figure}
\centerline{\epsfile{file=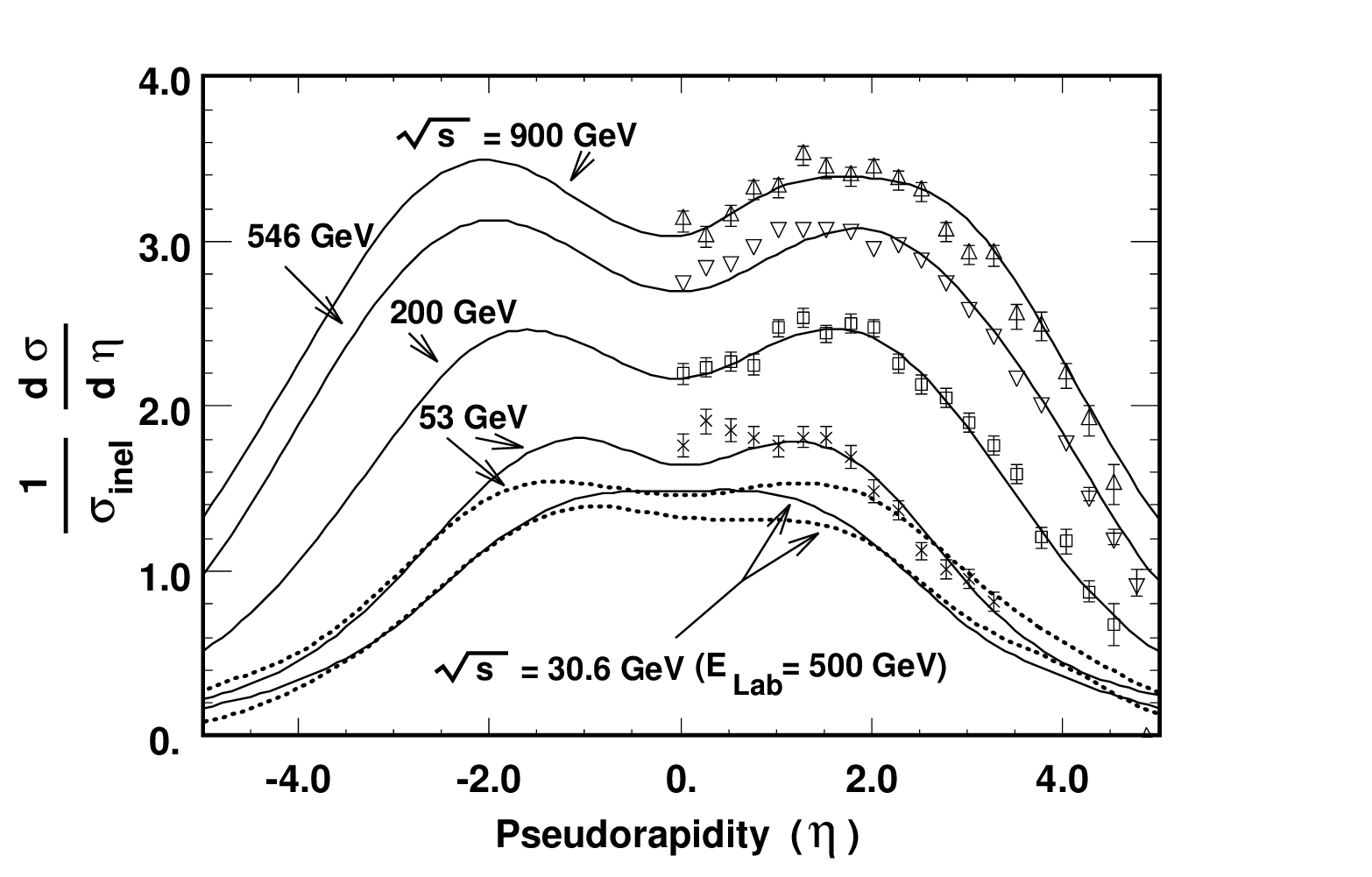,height=6.5cm}}
\caption{}
\vspace{5mm}
{\baselineskip=12pt
Pseudo-rapidity distributions calculated by the COSMOS code 
(solid line).
Upper triangles, downward triangles, and squares represent 
the data from UA-5, Sp${\rm \bar p}$S (Ref.~[52]) for 
$\sqrt{s}=$ 200, 546, 
and 900 GeV respectively.
Crosses are data from UA-5, ISR (Ref.~[53]) for 
$\sqrt{s}=$ 53 GeV.
The dotted lines show the calculated results by LUND code 
for $\sqrt{s}=$ 30.6 and 53 GeV.
}
\label{pp-eta}
\end{figure}

For high energies,
it is difficult to get experimental results of
nucleus--nucleus or proton--nucleus interactions.
However, 
there are many results available from 
p--${\rm\bar p}$(p) collider experiments.
In the Fig.~\ref{pp-eta}, we 
present the experimentally determined
pseudo-rapidity distribution for $\sqrt{s}=$ 53, 200, 546, 
and 900~GeV, and calculated results from the COSMOS and LUND 
codes for the same energies and $\sqrt{s}=$ 30.6~GeV,
corresponding to lab. energy of 500~GeV.
Above $\sqrt{s}=$ 53~GeV, the results of the COSMOS code agrees with
the experimental results within $\lesssim$ 5~\%.
Also the agreement of the COSMOS and LUND codes is good at
$\sqrt{s}=$ 30.6~GeV.
However, 
the pseudo-rapidity distributions calculated by the LUND code
is lower than the experiment results and those of the COSMOS 
code near $\eta=0$.
Accordingly the multiplicity (the number of particles created by the 
interaction) in the LUND code is smaller than
the experimental value above this energy.
Therefore, above 500 GeV in lab. energy we used the COSMOS 
code for hadronic interactions.

In the COSMOS code, 
nucleus -- nucleus interactions are treated as follows;
first, 
the projectile (cosmic ray) nucleus fragments into smaller 
mass number nuclei and nucleons with given probabilities.
Second, each fragmented nucleon interacts with 
the target independently, and the sum of created particles 
in each interaction is considered as the product of 
the nucleus -- nucleus interaction.
We take the 
average number of nucleons which interact with the target 
(air) nucleus as;
\begin{equation}
<N>= A \cdot {\sigma_p^{inel}\over \sigma_A^{inel}}
\label{eq-int-number}
\end{equation}
where, $A$ is the mass number of projectile nucleus, 
$\sigma_p^{inel}$ ($\sigma_A^{inel}$)
is the inelastic cross-section of the proton (nucleus) 
with the air nucleus.
We used the full treatment of COSMOS for the calculation 
of $\nu$--fluxes in the 30~MeV -- 3~GeV range,
and the superposition model in the 1 -- 3,000~GeV range.
The validity of the superposition model is discussed later 
(Sec.~\ref{sec:nflx-high}).

A qualitative discussion of the hadronic interactions is 
useful in order to understand the $\nu_\mu$ and $\nu_e$ 
excess over their antiparticles.
In hadronic interactions of low energy 
where one $\pi$ production is dominant,
the projectile charge is most often carried by the 
leading particle.
Thus, the most probable interaction for cosmic ray 
protons of this energy is;
\begin{equation}
\begin{array}{clcr}
p + A_{air} \rightarrow  &n + \pi^+ + X\\
\end{array}\label{eq:pi+}
\end{equation}
where $p$ stands for a proton and $n$ for a neutron.
In the higher energy region where protons 
cause the multiple production, there also is a similar 
effect that the $\pi^+$'s produced in this interaction 
statistically have a larger energy than $\pi^-$'s 
(see Fig.~\ref{p24be}).
Integrating with the steep spectrum of 
cosmic rays, there is an excess of $\pi^+$'s over $\pi^-$'s 
by $ \sim$ 20 \% for $\pi$--decay.
Consequently, we expect $\sim$ 20\% excess of 
$\nu_e$ over $\bar \nu_e$.
We note the neutron component of cosmic rays,
which is carried by helium and other nuclei,
creates a $\pi^-$'s excess over $\pi^+$'s from 
isospin--symmetry.
However, this gives a small effect on the $\pi$--spectrum and 
the atmospheric neutrino flux due to the proton dominance 
in the cosmic ray flux.

\subsection{Decay of Mesons}
\label{sec:decay}

Neutrinos are mainly produced in the following decays of $\mu$'s, 
$\pi$'s, and $K$'s~\cite{particle-data}:
\begin{equation}
\begin{array}{clcrc}
  \pi^\pm  &  \rightarrow \mu^\pm \nu_\mu(\bar \nu_\mu) &  & (100~\%)&\\
  \mu^\pm  &  \rightarrow 
e^\pm \nu_e(\bar \nu_e) \bar \nu_\mu (\nu_\mu) &  & (100~\%)&\\
  K^\pm   &  \rightarrow \mu^\pm \nu_\mu(\bar \nu_\mu)&  & (63.5~\%)&\\
       &  \rightarrow \pi^\pm \pi^0 &  & (21.2~\%)&\\
       &  \rightarrow \pi^\pm \pi^+ \pi^- &  & (5.6~\%)&\\
       &  \rightarrow \pi^0 \mu^\pm \nu_\mu(\bar \nu_\mu) &  
& (3.2~\%)& (K_{3\mu\nu})\\
       &  \rightarrow \pi^0 e^\pm \nu_e(\bar \nu_e) &  
& (4.8~\%)& (K_{3e\nu}) \\
       &  \rightarrow \pi^\pm \pi^0 \pi^0 &  & (1.73~\%)&\\
\\
 K^0_s  &  \rightarrow \pi^+ \pi^-  & & (68.6~\%)&\\
\\
 K^0_l  &  \rightarrow \pi^+ \pi^- \pi^0  & & (12.37~\%)&\\
       &  \rightarrow \pi^\pm \mu^\mp \nu_\mu(\bar \nu_\mu) &  & (27~\%)
& (K_{3\mu\nu})\\
       &  \rightarrow \pi^\pm e^\mp \nu_e(\bar \nu_e)  & & (38.6~\%)
& (K_{3e\nu})\\
\end{array}
\end{equation}
The decay of charged $\pi$'s and subsequent $\mu$ decay 
($\pi-\mu$ decay):
\begin{equation}
\begin{array}{clcr}
\pi^\pm  \rightarrow  &\mu^\pm + \nu_\mu (\bar \nu)_\mu\\
              & \downarrow \\
 &\mu^\pm \rightarrow 
e^\pm + \nu_e(\bar \nu)_e + \bar \nu_\mu (\nu_\mu),\\
\end{array}\label{eq:pi-mu}
\end{equation}
is dominant among these processes. 
Charmed particles, such as $D$ and $\bar D$
also create $\nu$'s, however,
the contribution of charmed particle to atmospheric $\nu$'s
becomes sizable only for $E_\nu \gtrsim 100$~TeV, 
which is far beyond the energy region of concern here.

When a $\pi^\pm$ decays at rest,
the energy carried by $\nu_\mu (\bar \nu_\mu)$ is 
$(m_\pi^2 - m_\mu^2)/2m_\pi 
\sim$ 30~MeV, and the $\mu^\pm$ carries the rest of the energy.
If we ignore the spin of the $\mu$'s,
each decay particle, $e^\pm$,
$\nu_e (\bar \nu_e)$, and $\bar \nu_\mu (\nu_\mu)$
carries $1/3$ of $\mu^\pm$'s energy ($\sim 37$~MeV) 
on average in the three body decay.
When $\pi$'s decay in flight,
the $\pi$ energy is approximately divided into $1/4$ 
to each decay product in the $\pi-\mu$ decay on average.
Thus, we expect the flux ratio to be  roughly 
$(\nu_e + \bar \nu_e)/ (\nu_\mu + \bar \nu_\mu) = 1/2$
and $\bar \nu_\mu/\nu_\mu = 1$ irrespective of the $\pi$ spectrum.
When the energy of the $\mu$'s becomes high ($\gtrsim 5$~GeV), 
however,
$\mu$'s tend not to decay in the air but
to reach the Earth before decaying.
In this case, $\mu$'s lose their energy in the Earth, 
and decay after they are almost stopped,
or are captured by nuclei in the Earth.
This effect reduces the ratio
$(\nu_e + \bar \nu_e)/ (\nu_\mu + \bar \nu_\mu)$
above this energy.
Also the ratio $\bar\nu_\mu/\nu_\mu$ decreases with 
energy in the same energy region, 
since cosmic ray protons create 
more $\pi^+$'s than $\pi^-$'s, and 
there is a corresponding excess of $\mu^+$'s over $\mu^-$'s.
We note that
$\mu$ polarization and $\mu$ energy loss in air
are important for the precise calculation of 
atmospheric $\nu$'s.
Each has $\gtrsim$ 5~\% effects on the $\nu$'s energy created by 
$\mu$'s.
The energy loss of $\mu$'s in the air is taken into account 
by the Monte Carlo method.
The treatment of $\mu$ polarization is explained in the following.

In the decay of charged $\pi$'s,
the resulting $\mu^\pm$ is fully polarized 
against (toward) the direction of $\mu$ 
motion in the charged $\pi$ or $K$ rest frame.
In the subsequent $\mu$ decay, 
$\nu_e(\bar \nu_e)$ is emitted to the forward direction of 
the $\mu$'s motion from the conservation of helicity.
Thus the $\nu_e(\bar \nu_e)$ resulting from $\mu$ decay 
has a larger energy than $\bar \nu_\mu (\nu_\mu)$
in the $\pi$ rest frame.
Since $\pi$'s decay in flight,
the polarization is not full.
In general, 
the direction distribution of $\nu_e$ is proportional to 
$(1+\zeta\cos\theta)$
where $\zeta$ is the polarization
($\zeta=1$ is full polarization) and
$\theta$ is the angle between
directions of $\nu_e$--motion and $\mu$--spin
in the CM frame of the $\mu$'s.
Practically, it is the polarization in the
observer's frame that is important.
Using the spin direction 3-vector ${\vec \zeta}$,
the polarization along the $\mu$ momentum direction
is 
\begin{equation}
\zeta = \vec \zeta \cdot {{\vec p}\over |{\vec p}|} 
= {E E^* - \gamma_\pi m_\mu^2 \over {\vec p}\cdot {\vec p}^* } ,
\label{eq:mupol}
\end{equation}
in the observer's frame~\cite{hayakawa}.
Here $E$ is the $\mu$ energy, ${\vec p}$ the $\mu$ momentum ,
and $\gamma_\pi$ is the Lorentz factor of the $\pi$ 
in observer's frame.
$E^*$ and ${\vec p}^*$ denote the $\mu$ energy and the momentum 
in the $\pi$ rest frame respectively.

The above discussion can also be applied to $\mu$'s created
in the  $ K^\pm  \rightarrow \mu^\pm \nu_\mu(\bar \nu_\mu)$
decay.
For $\mu$'s resulting from the $K_{3l\nu}$ decay, the discussion in 
Ref. \cite{brene} is applied.
The small angle scattering of $\mu$'s in the atmosphere
reduces the $\mu$ polarization.
This depolarization effect is also evaluated by 
Hayakawa~\cite{hayakawa}
as of the order of $21{\rm MeV}/v p$, where $v$ and $p$ are
velocity and momentum of $\mu$'s respectively.
Therefore, the depolarization effect may be negligible for 
$\mu$'s which produce $\nu$'s of energies $\gtrsim 100$~MeV.
If we ignore the effect of $\mu$ polarization, 
the calculated energy of $\nu_e$ decreases by $\sim 5$~\% on average,
therefore, the flux of $\nu_e$ is estimated to be smaller by 
$\sim$ 10~\% at 500~MeV and $\sim$ 15\% at 3~GeV.
This is an important effect for low energy $\nu$'s,
which are observed in underground detectors~\cite{volkova2}.

\section{The Flux of Atmospheric neutrinos}
\label{sec:nflx}

\subsection{$\nu$--fluxes and Atmospheric Density Structure}
\label{sec:atmosphere}

We note first that the density structure of the atmosphere is 
important, 
because it is the reason for a large zenith angle dependence
of $\nu$--fluxes.
We take the US standard~\cite{uss1} for the density structure of 
the atmosphere.
The chemical composition of atmosphere is approximated by
nitrogen 78~\%, oxygen 21~\%, and argon 1~\% in our calculation. 
For $\pi$'s and $K$'s propagating in the atmosphere,
the decay and interaction with air nuclei are competitive processes.
When the relation
\begin{equation}
c \tau {E \over mc^2} \sim { 1 \over \sigma n}
\label{eq:pint1}
\end{equation}
is satisfied, both processes work at nearly the same rate. 
Here, $\tau$ is the life time of the meson,
$E$ is the energy, 
$\sigma$ the interaction cross--section of 
the meson and air nuclei,
and $n$ the number density of air nuclei.
This condition~(\ref{eq:pint1}) is rewritten for the energy
of mesons with which decay and interaction take place almost equally
as
\begin{equation}
E \sim {mc^2 \over c\tau\sigma n} =
\left\{
\begin{array} {cl}
12\ ({\rm GeV,\ for}\ \pi^{\pm})\\
22\ ({\rm GeV,\ for}\ K_L^0)\\ 
90\ ({\rm GeV,\ for}\ K^{\pm})\\
\end{array}\right\}
 \times {\rho_{[sea\ level]} \over \rho}\ ,
\label{eq:pint2}
\end{equation}
where $\rho_{[sea\ level]} = 1.225 {\rm kg m^{-3}}$. 
Most mesons which have smaller energies than that given
by this equation decay.
Slant entering cosmic rays interact with atmospheric nuclei
at a higher altitude than 
vertically entering cosmic rays on average.
Therefore, the decay probability is larger for the mesons created by  
slant entering cosmic rays than for those of 
vertical cosmic rays.
We expect a larger $\nu$--flux from near horizontal directions 
than from the vertical.

The first interaction of vertical cosmic rays takes place 
at an altitude of $15 \sim$ 20 km, 
where the density of air is $\sim$ 10 times less than 
that at the sea level,
since the interaction mean-free-path (MFP) for cosmic ray protons is 
$\sim 100\ {\rm g/cm^2}$ in column density.
Mesons with energy $\lesssim 12 $~GeV,
which create $\nu$'s of $\lesssim 3$~GeV,
decay before interacting with  air nuclei.
However, for mesons with energy $\gtrsim 100$~GeV, 
the probability of interaction becomes sizable.
Therefore, the variation of atmospheric $\nu$--fluxes 
with the zenith angle increases with the $\nu$--energy.

\subsection{Low Energy $\nu$ Flux (30~MeV -- 3~GeV)}
\label{sec:nflx-low}

At low energies, although the zenith angle dependence of 
$\nu$--fluxes caused by the atmospheric structure is not large,
a significant directional variation is caused by the rigidity cutoff.
In the one--dimensional approximation we adopted,
we expect larger $\nu$--fluxes from the low rigidity cutoff 
direction and a smaller $\nu$--fluxes from the high rigidity cutoff
direction.
There should be a large directional dependent variation of $\nu$--fluxes
for Kamioka (Fig.~\ref{kamcut}), especially a large deficit 
from east--horizontal direction.
In the actual case, however, it may be difficult 
to observe those variations.
The direction of a $\nu$ is different from that of the parent cosmic ray,
because
the mesons are produced with slightly different directions
from that of incident particle,
and again the $\nu$'s are created with slightly different
directions from those of the mesons in their decay.
We expect this effect to be rather small, but there is another
smearing effect of direction in the $\nu$--detector.
When a low energy $\nu$ ($\lesssim 3$~GeV)
creates a lepton by a quasi-elastic process,
the lepton has a typical angle of $50 - 60^\circ$
from the $\nu$ direction~\cite{takita}.
Therefore, it may be difficult to 
observe the deficit of $\nu$--flux from 
east--horizontal direction in Kamioka.
Thus the direction dependence of atmospheric $\nu$--flux 
is small for lower energy $\nu$'s, especially when they are
observed in the detector.
We present here the atmospheric $\nu$--flux 
averaging over all directions. 

For low energy $\nu$--fluxes, we employed a full Monte Carlo method.
The calculation itself is rather straightforward;
first, the nucleus and the primary energy
of the cosmic ray are sampled with equation~(\ref{eq:1ryeq1}).
Second, the arrival direction is sampled uniformly.
When the rigidity is smaller than the rigidity cutoff,
the cosmic ray is discarded.
When the rigidity is larger than the rigidity cutoff, 
the cosmic ray is put to the propagation code of 
cosmic rays in the atmosphere,
which controls the interaction of cosmic rays,
the decay of secondary particles, and the energy losses in the 
atmosphere.
The COSMOS code controls all these processes.

\begin{figure}
\centerline{\epsfile{file=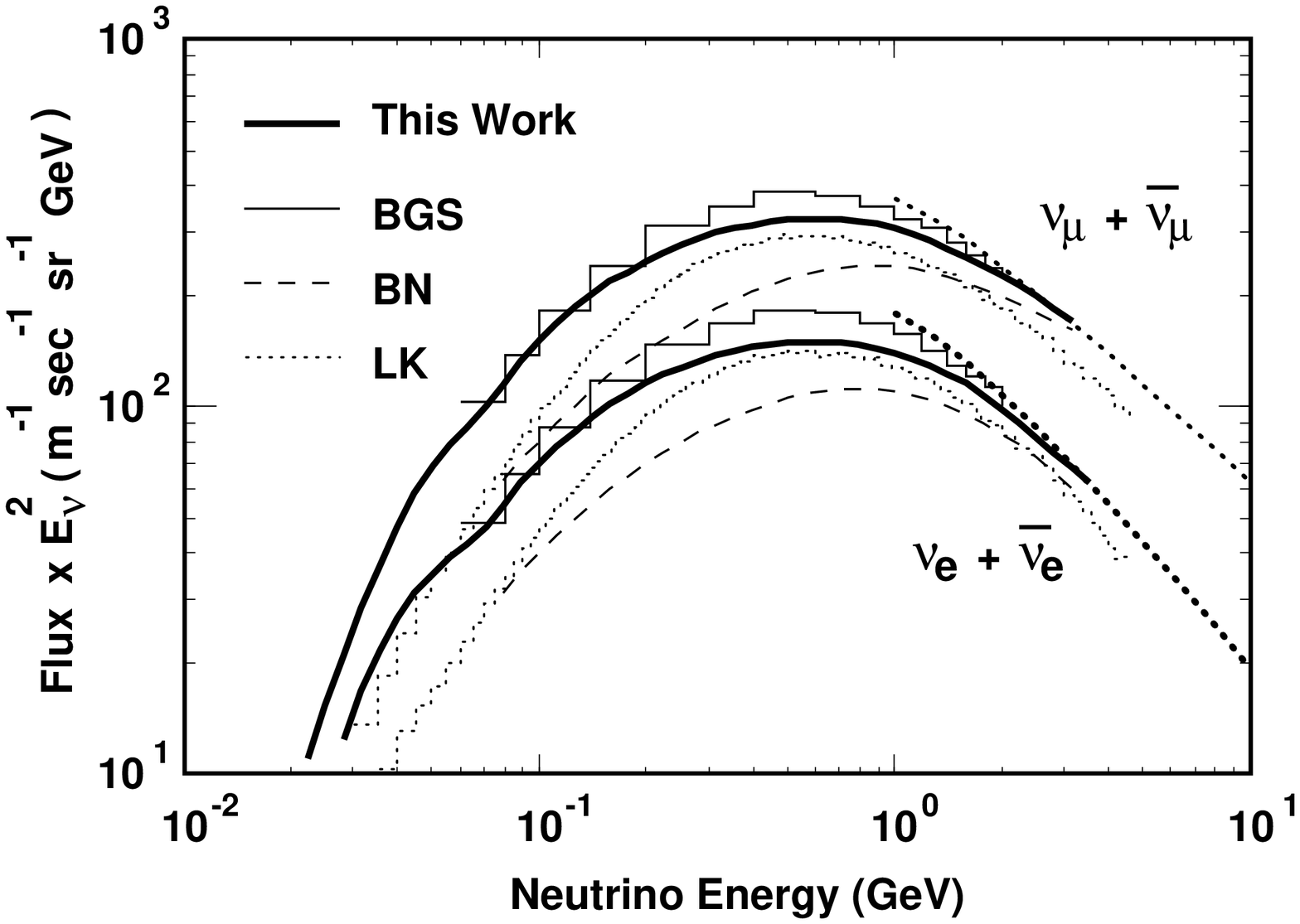,height=10cm}}
\caption{}
\vspace{5mm}
{\baselineskip=12pt
The atmospheric $\nu$--fluxes multiplied by $E_\nu^{2}$ 
for the Kamioka site at solar mid. (solid lines). 
BGS are from Ref.~[16], %[bar89] 
BN from Ref.~[17], %[bn89]
and LK from Ref.~[18]. %[lk90]
The dotted line is the result from the calculation for high 
energy without the rigidity cutoff, and averaged over all 
directions. 
For details, see the next section (IV C).
}
\label{fig:kamnflx}
\end{figure}

The results are summarized in Table~\ref{low-nu-kam} for Kamioka
and in Table~\ref{low-nu-kam} for the IMB site, both for solar mid.
Also in Fig.~\ref{fig:kamnflx},  $\nu_\mu + \bar \nu_\mu$ and
$\nu_e + \bar \nu_e$ fluxes are 
depicted for Kamioka at solar mid., and compared with the other results.
Since $\nu$--fluxes for solar mid. are not available for other 
authors, we averaged solar max. and solar min. values.
Flux differences between solar max. and solar min. are
$\sim$ 8~\% at 100~MeV and $\sim 3$~\% at 1~GeV for Kamioka,
and $\sim$ 12~\% at 100~MeV and $\sim 4$~\% at 1~GeV
for IMB due to its lower cutoff-rigidity.
We note that these calculated results are smoothly 
connected to the atmospheric neutrino fluxes calculated by the
hybrid method for high energies (1 -- 3,000~GeV) at around 3~GeV,
where the effect of the rigidity cutoff is small.

\begin{figure}
\centerline{\epsfile{file=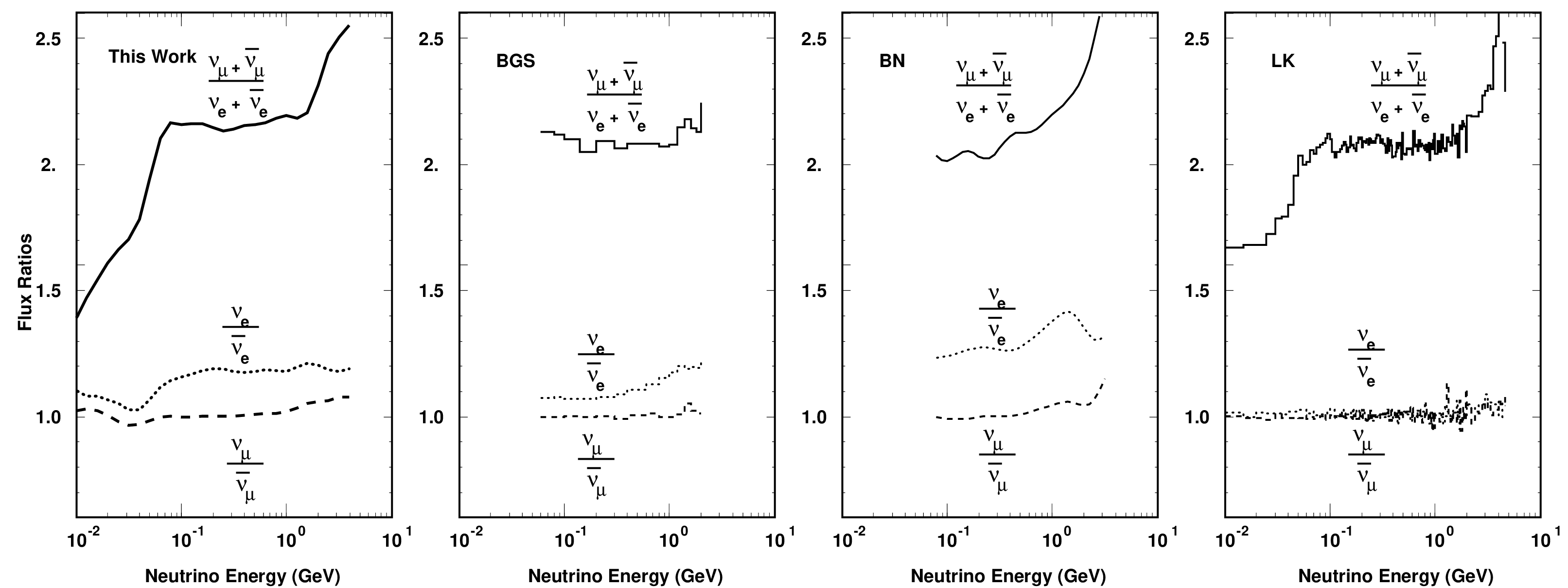,height=6.5cm}}
\caption{}
\vspace{5mm}
{\baselineskip=12pt
The flux ratio of $\nu$--species calculated for Kamioka.
BGS are from Ref.~[16], %[bar89]
BN from Ref.~[17], %[bn89]
and LK from Ref.~[18]. %[lk90]
as before. 
}
\label{fig:nflx-ratio}
\end{figure}

In Fig.~\ref{fig:nflx-ratio}, we show the flux ratio by $\nu$--species
along with those of other authors.
We note that although the calculation method and some of 
physical assumptions are different among these authors,
the ratio $(\nu_e + \bar\nu_e)/(\nu_\mu + \bar\nu_\mu)$ is 
very similar each other.
The relatively large difference in $\bar\nu_e/\nu_e$ among them may 
reflect the difference of calculation scheme and/or 
the physical assumptions.

\subsection{High Energy $\nu$ Flux (1 -- 3,000~GeV)}
\label{sec:nflx-high}

For the calculation of atmospheric $\nu$'s at high energies 
(1 -- 3,000~GeV), 
we employed the superposition model and a hybrid method.
We note that for helium nuclei, the complete treatment of COSMOS and 
the superposition model give essentially the same result.
Let us measure the atmospheric depth by the column density from the top 
of the atmosphere.
With equation~(\ref{eq-int-number}), the average number of nucleons 
which interact with an air nucleus in a depth interval $[x, x+dx]$ 
is given by 
\begin{equation}
<N> e^{-{x\over \lambda_A}} {dx\over \lambda_A} 
=A \cdot {\sigma_p \over \sigma_A} 
\cdot e^{-{x\over \lambda_A}}\ {dx\over \lambda_A} 
=A  \cdot {\lambda_A \over \lambda_p} \cdot
e^{-{x\over \lambda_A}}\ {dx\over \lambda_A} 
\hskip 1cm 
\label{eq-1st-int}
\end{equation}
per nucleus.
The nucleus MFP in the column density
of air is denoted by $\lambda_A$,
and the proton MFP denoted by $\lambda_p$.
We note that $\sigma \propto 1/\lambda$.
Since the fragmentation of helium nuclei into deuterons occur 
with only a small probability,
we can safely assume that 
helium nuclei always fragment into four nucleons \{$p,p, n,n$\}.
The non-interacting nucleons at the first collision also interact with
air nuclei in succeeding processes.
The average number of nucleons which did not interact in the helium 
collision at the depth $x_1$ and that do interact in  
$[x, x+dx]$ is given by 
\begin{equation}
(A-<N>)[\int_0^{x} {1\over \lambda_A} e^{-{x_1\over \lambda_A}}\cdot 
{1\over \lambda_p} e^{-{x-x_1\over \lambda_p}} dx_1]\ dx
= A \cdot [e^{-{x \over \lambda_p}} 
- e^{-{x\over \lambda_A}}]\ {dx \over \lambda_p}
\label{eq-2nd-int}
\end{equation}
for a nucleus.
Adding (\ref{eq-1st-int}) and (\ref{eq-2nd-int}), 
we get the average number of interacting nucleons in a depth
interval $[x, x+dx]$ as 
\begin{equation}
A \cdot e^{-{x\over \lambda_A}}\ {dx \over \lambda_p}
+ A \cdot
[e^{-{x \over \lambda_p}}- e^{-{x\over \lambda_A}}]\ {dx \over \lambda_p}
= A e^{-{x\over \lambda_p}}\ {dx \over \lambda_p}, 
\end{equation}
which is exactly the same with the case of $A$-independent nucleons.
We note that up to $E_{cr}=100$~TeV, 
around 80~\% of all nucleons are carried by hydrogen and helium nuclei.

For heavier cosmic rays than helium, the probability of
fragmentation into smaller nuclei 
(e.g., helium nucleus = $\alpha$ particle, etc.) is 
large at collisions.
Therefore, 
the average number of nucleons which interact with the target (air) nucleus
is smaller than that given by equation~(\ref{eq-int-number}).
Thus, the interaction height (depth) distribution of each nucleon 
is different from proton cosmic rays in general.
This distribution is represented by the effective MFP of 
nucleon for heavier nuclei,
and is calculated by the COSMOS code as 
$\sim 100~{\rm g/cm^2}$ for CNO and $\sim 94~{\rm g/cm^2}$ 
for Fe nuclei.
The MFP is $ 87.4~{\rm g/cm^2}$ for protons,
$36.6~{\rm g/cm^2}$ for CNO's, and $15.6~{\rm g/cm^2}$ 
for Fe nuclei at 1~TeV$/$nucleon.
Thus the effective MFP for nucleons carried by those nuclei
is more similar to the proton MFP than to those nuclei.
These facts reasonably justify the superposition model,
even for nuclei heavier than heliums.

In our hybrid method,
we calculate the $\nu$--yield function for protons,
$\eta^p_{\nu}(E, E_\nu, \theta)$,
which denotes the number of $\nu$ in the energy region 
from $E_\nu$ to $E_\nu + dE_\nu$, created by 
a proton with energy of $E$ incident from zenith angle
$\theta$ with the Monte Carlo method.
We executed the Monte Carlo simulation for 
proton cosmic rays with energies  
from 1~GeV to 1,000~TeV in $\Delta \log(E_p)= 0.2$ steps,
and for zenith angle bins of 
$\cos \theta =$ 0.0 -- 0.1, 0.1 -- 0.2, .... 0.9 -- 1.0.
For each zenith angle bin, 300,000 protons were injected to the
Monte Carlo code at 1~GeV, 
100,000 protons at 10~GeV,  
....,
1,000 protons at 100~TeV,  
and 300 protons at 1,000~TeV.

\begin{figure}
\centerline{\epsfile{file=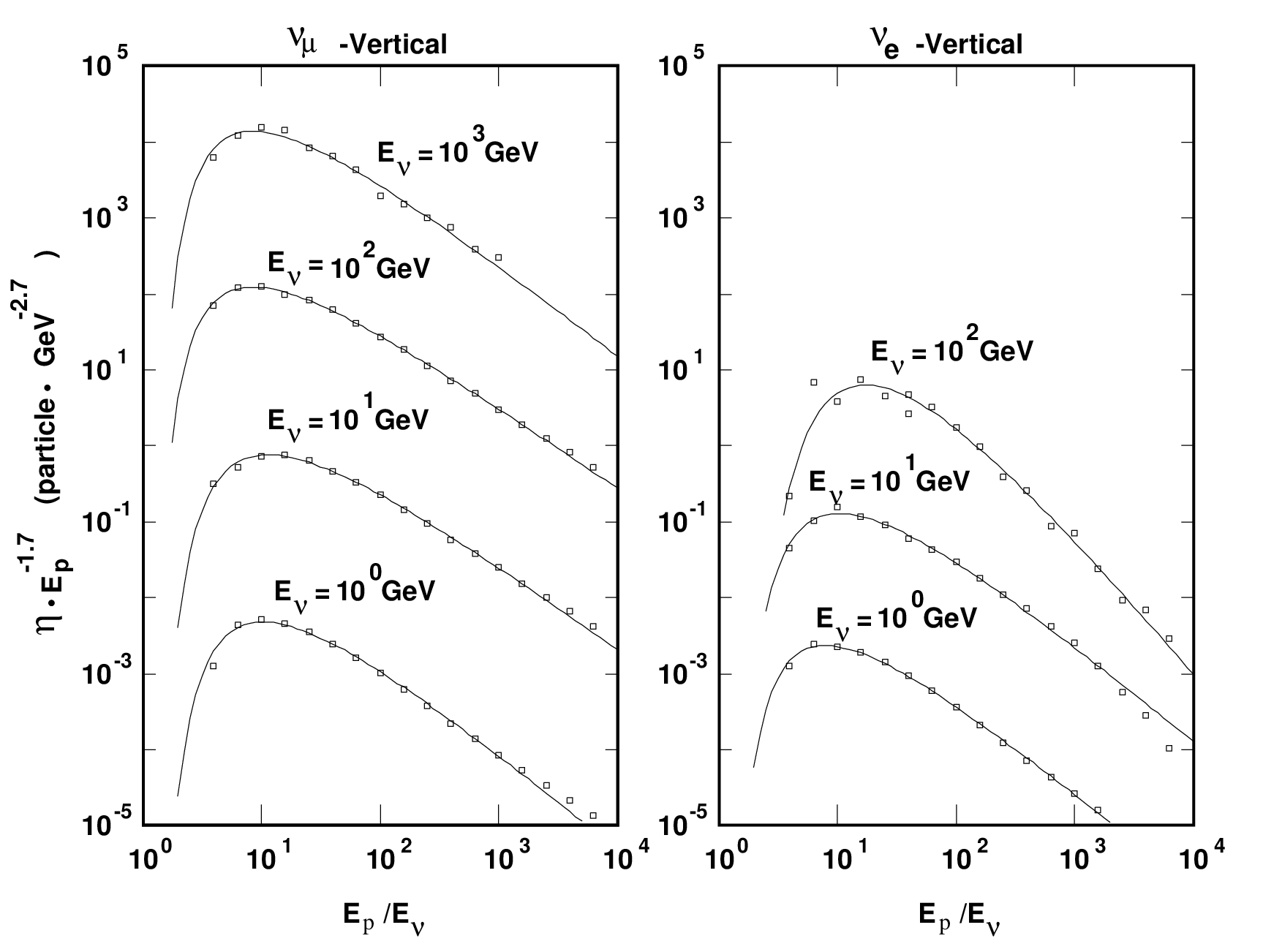,height=10cm}}
\caption{}
\vspace{5mm}
{\baselineskip=12pt
$\nu$--yield function for protons with fixed $E_\nu$ 
multiplied by $ E_p^{-1.7}$
for $\nu_\mu$ (left) and $\nu_e$ (right) for near vertical directions
($\cos \theta = 0.9 - 1.0$).
Monte Carlo results are shown by squares,
and fitted results by lines.
}
\label{fig:checkfit}
\end{figure}

%%%%%%%%%%%
The proton $\nu$--yield function for $\nu_\mu$'s and $\nu_e$'s
calculated by the Monte Carlo method
is fitted by a function:
\begin{equation}
F(E_p, E_\nu)
= (E_p)^{-1.7} \times 
A\cdot {(\log(E_p/E_\nu) - B)^2 \over \log(E_p/ E_\nu)} + C \ ,
\label{eq:fitfun}
\end{equation}
with parameters $\{A,B,C\}$.
This fit agrees with the result of the Monte Carlo method
very well in the region $3 \lesssim E_p/E_\nu \lesssim 10^3$,
as is seen in Fig.~\ref{fig:checkfit}.
The solid lines show our fit with equation \ref{eq:fitfun}.
As the cosmic ray spectrum is approximately proportional to 
$E_{cr}^{-2.7}$ in a wide energy range,
the quantity shown in  Fig.~\ref{fig:checkfit}
roughly stands for  
the relative contribution to a fixed energy $\nu$'s 
from the cosmic ray in a logarithmic energy bin.
The largest contribution 
to fixed energy atmospheric $\nu$'s 
comes from the cosmic ray with $E_{cr} \sim 10 E_\nu$
both for $\nu_\mu$'s and $\nu_e$'s.
We note that the contribution from $E_p/E_\nu \lesssim 3$
and $E_p/E_\nu \gtrsim 10^3$ is very small($\lesssim 5$~\%). 
The yield functions of $\nu_e$'s and $\bar\nu_e$'s 
decreases more rapidly than $\nu_\mu$'s and $\bar\nu_\mu$'s,
and they change their shape for $E_\nu \gtrsim 100$~GeV.

%%%%
For the neutron incident,
we assumed 
\begin{equation}\begin{array}{l}
\eta^p_{\bar \nu}(E, E_\nu, \theta) =\eta^n_{\nu}(E, E_\nu, \theta)\
\ \ {\rm and}\\
\eta^p_{\nu}(E, E_\nu, \theta) =\eta^n_{\bar \nu}(E, E_\nu, \theta)\ .\\  
\label{eq:isospin--symmetry}
\end{array}\end{equation}
This assumption is justified for $\nu$'s produced through
$\pi$'s,
but not for those produced through $K$'s.
The $K^-/K^+$ ratio is rather an universal quantity for p--A and 
A--A interactins ($\sim 0.8$) at high energies~\cite{akesson}.
However, we note that the portion of proton nucleon is still $\sim 80$~\%
in the cosmic ray at 100~TeV (Fig.~\ref{fig:1ry-ratio}).
Therefore, the assumption
leads to a maximum of 10~\% errors in the $K^-/K^+$--ratio.
We note that this assumption affects only on the 
$\nu/\bar\nu$ ratio at high energies.
We expect almost no effect for $\nu/\bar\nu$ ratio at low energies, 
because of the proton dominance in the low energy cosmic ray.
Also the $\nu +\bar\nu$--flux is not affected in any energy region.

The atmospheric $\nu$--fluxes were calculated by integrating 
those $\nu$--yield functions with the nucleon fluxes shown
in Fig~\ref{fig:nucleon}:
\begin{equation}
F_{\nu}(E_\nu) = \int_{Emin}^{Emax} 
[\eta^p_{\nu} (E, E_\nu, \theta) \times F_p(E)
+ \eta^p_{\bar \nu}(E, E_\nu, \theta) \times F_n(E)] dE
\label{eq:flxint1}
\end{equation}
and
\begin{equation}
F_{\bar\nu}(E_\nu) = \int_{Emin}^{Emax} 
[\eta^p_{\bar\nu}(E, E_\nu, \theta) \times F_p(E)
+ \eta^p_{\nu}(E, E_\nu, \theta) \times F_n(E)] dE\ ,
\label{eq:flxint2}
\end{equation}
where $F_\nu(F_{\bar\nu})$ is the atmospheric 
$\nu(\bar\nu)$--flux,
$F_p$ and 
$F_n$ are 
the proton and neutron--fluxes respectively,
and $\nu$ stands for $\nu_\mu$ or $\nu_e$.
We took $E_{min}$ equal to the $\nu$--energy and
$E_{max}=1,000$~TeV. 

%%%%%%%% maximum and special values at $1, 3.16\times 10^4$~GeV
The results of the integrations (\ref{eq:flxint1}) and 
(\ref{eq:flxint2}) are summarized in Tables~\ref{s-numu3},
\ref{s-numu-bar3},\ref{s-nue3}, and \ref{s-nue-bar3} down to 1~GeV.
The calculated fluxes below 10~GeV smoothly 
connect to the low energy calculation at around 3~GeV
(Fig.~\ref{fig:kamnflx}).
This shows that the systematic difference between the calculations for
high energy and low energy atmospheric $\nu$--fluxes
(section~\ref{sec:nflx-low}) is very small.
With the limited number of protons used in the calculation of 
$\nu$--yield function at high energies,
the errors involved in $\{A,B,C\}$ become large due to large fluctuations.
Therefore, the error on the flux value increases at high energies.
This error is estimated to be $\lesssim 5$~\% for $\nu_\mu$ and 
$\bar\nu_\mu$ below 100~GeV,
and increases to 10~\% at around $3,000$~GeV for 
near horizontal directions.
The error for $\nu_e$ and $\bar\nu_e$ fluxes is larger than 
$\nu_\mu$ and $\bar \nu_\mu$ fluxes; it may exceed 10~\% at 
1,000~GeV for near horizontal directions and at 100~GeV for near 
vertical directions.
For $\nu_\mu$'s and $\bar \nu_\mu$'s, however,
we extrapolate the flux value up to $3.16\times 10^4$~GeV.
We note that large errors at these energies do not affect much the 
expected flux of up--going $\mu$'s, because the contribution
of $\nu$'s with $E_\nu > 1,000$~GeV is estimated less than 15~\%.
The value in the parentheses is estimated to have a larger error 
than 10~\%.

\begin{figure}
\centerline{
\epsfile{file=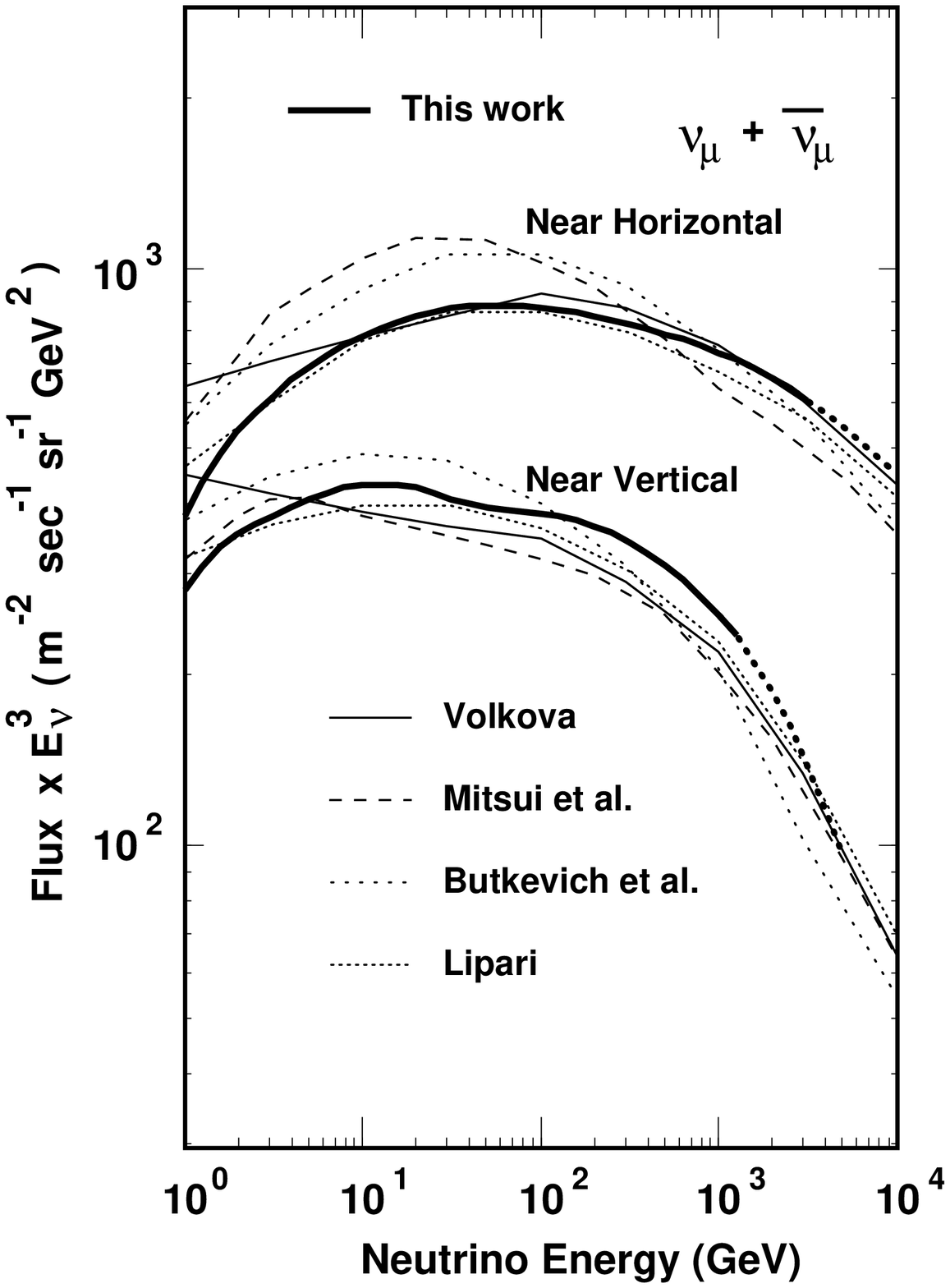,height=10cm}
\epsfile{file=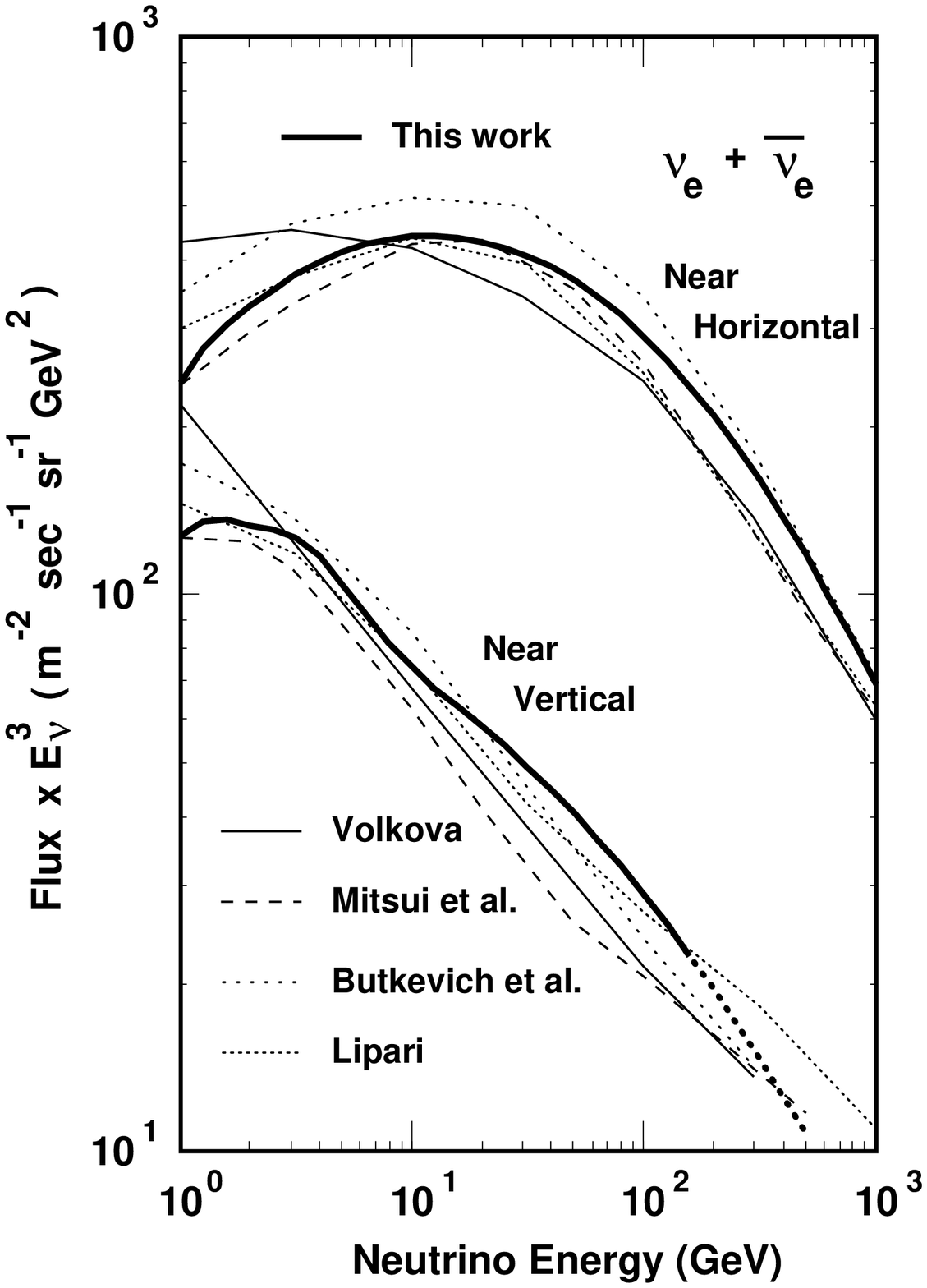,height=10cm}}
\caption{}
\vspace{5mm}
{\baselineskip=12pt
$\nu_\mu + \bar \nu_\mu$ and $\nu_e + \bar \nu_e$ fluxes 
for near horizontal ($\cos\theta = 0-0.1$) and 
near vertical ($\cos\theta = 0.9-1.0$) directions.
Volkova is from Ref.~[12], %[volkova] 
Mitsui et al from Ref.~[13], %[mitsui]
Butkevich et al from Ref.~[14], %[butkevich]
and Lipari from Ref.~[15]. %[lipari]
We used the fluxes of $\cos\theta = 0.05$ for near horizontal 
and $\cos\theta = 1$ for near vertical directions 
of Volkova and Lipari, 
those of $\theta_{zenith}=87^\circ$ for near horizontal and 
$\theta_{zenith}=0$ for near vertical directions of Mitsui et al, and
those of $\cos\theta = 0$ for near horizontal and $\cos\theta = 1$ 
for near vertical directions of Butkevich et al.
}
\label{fig:nue}
\end{figure}

The $\nu$--fluxes below 3~GeV could differ substantially 
from the true flux, since  solar 
modulation and geomagnetic effects have not been included.
The fluxes for $\le$ 3~GeV in these tables should be regarded 
as those for solar mid. and low cutoff rigidity ($\lesssim 3$~GV).
They also are depicted for the near vertical and near horizontal directions
in Fig.%s~\ref{fig:numu} and 
\ref{fig:nue} with the results of other 
authors.
The $\nu_\mu (\bar\nu_\mu)$ fluxes are almost 
proportional to $E_\nu^{-3}$ from 1~GeV to 1~TeV.
On the other hand,
the $\nu_e (\bar\nu_e)$ fluxes decrease, being proportional to
$E_\nu^{-3.5}$ or steeper above 10~GeV.
Our $\nu_\mu$--flux for near horizontal directions
agrees with that of Volkova~\cite{volkova} and Lipari~\cite{lipari} 
above 10~GeV to within $\sim$ 5~\%.
For near vertical directions, our calculation is 10 -- 15 \% larger than
that of Volkova and Mitsui et al.~\cite{mitsui}, 
and by 5 -- 10\% than that of Lipari.
For $\nu_e$--fluxes, our calculation is larger than others 
by 10-20\% above 10~GeV for both directions.

If we take the lower line for the extrapolation of 
the proton flux in Fig.~\ref{fig:nucleon} for $\ge 40$~TeV,
the $\nu_\mu$ and $\bar\nu_\mu$--fluxes decrease $\sim$ 10~\% at 3,000~GeV,
but they decrease only a few \% at 1,000~GeV.
Even if we reduce the upper end of the integrations ($E_{max}$)
to 100~TeV in (\ref{eq:flxint1}) and (\ref{eq:flxint2}),
i.e. if there was a sharp cutoff of the cosmic ray spectrum at 1,000~TeV,
the $\nu_\mu$ and $\bar \nu_\mu$ fluxes are reduced by only $\sim$10~\%
at 1,000~GeV.
Therefore, even if the cosmic ray composition above 100~TeV is quite 
different from our assumption,
our calculation does not give a different result
for the atmospheric $\nu$--fluxes below 1,000~GeV by more than 10~\%.
However, for the accurate calculation of atmospheric $\nu$'s 
above 1,000~GeV, it is necessary to determine the flux and composition of 
primary cosmic rays above 1,000~TeV per nucleon accurately.

\begin{figure}
\centerline{
\epsfile{file=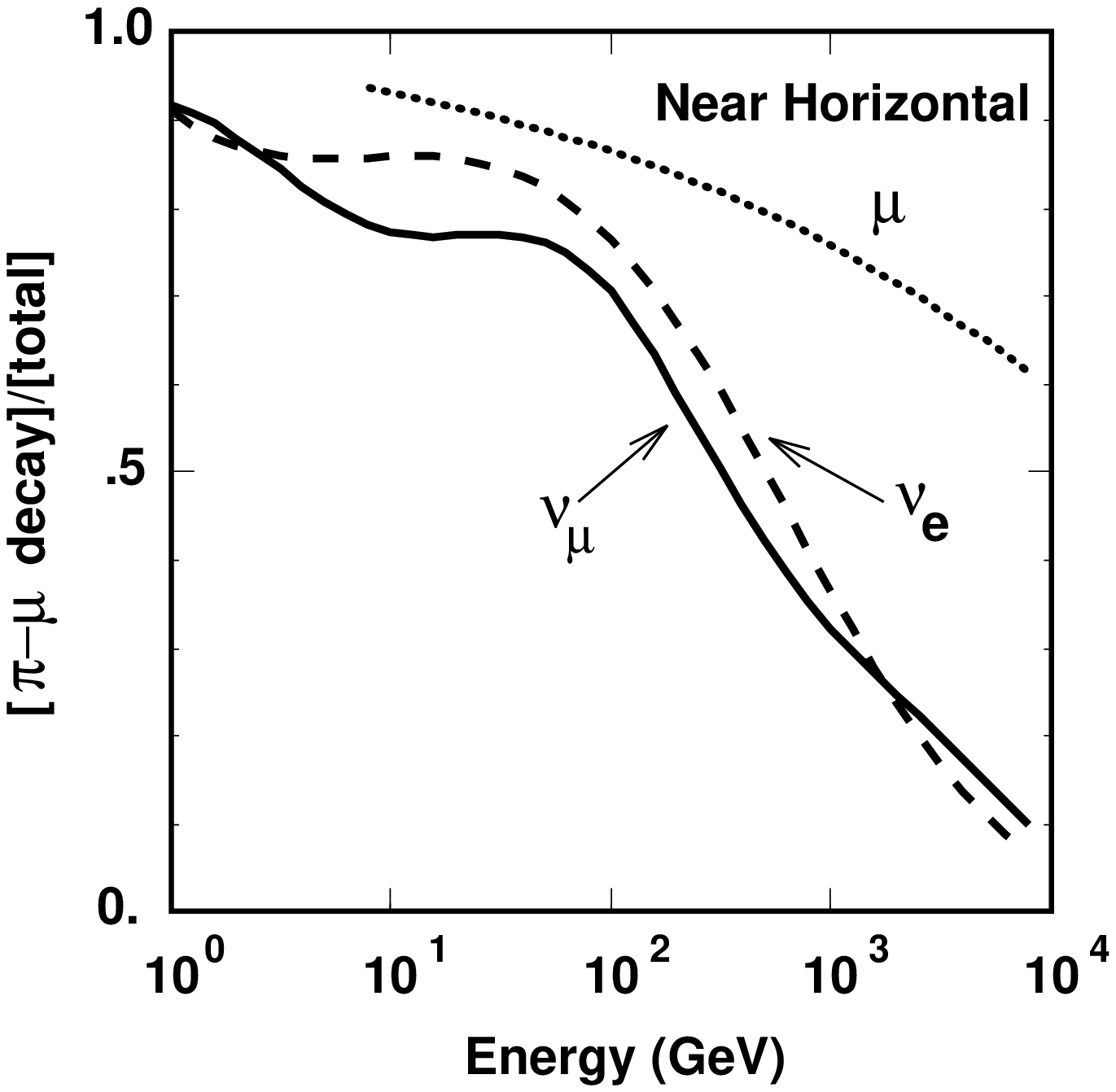,height=6.5cm}
\epsfile{file=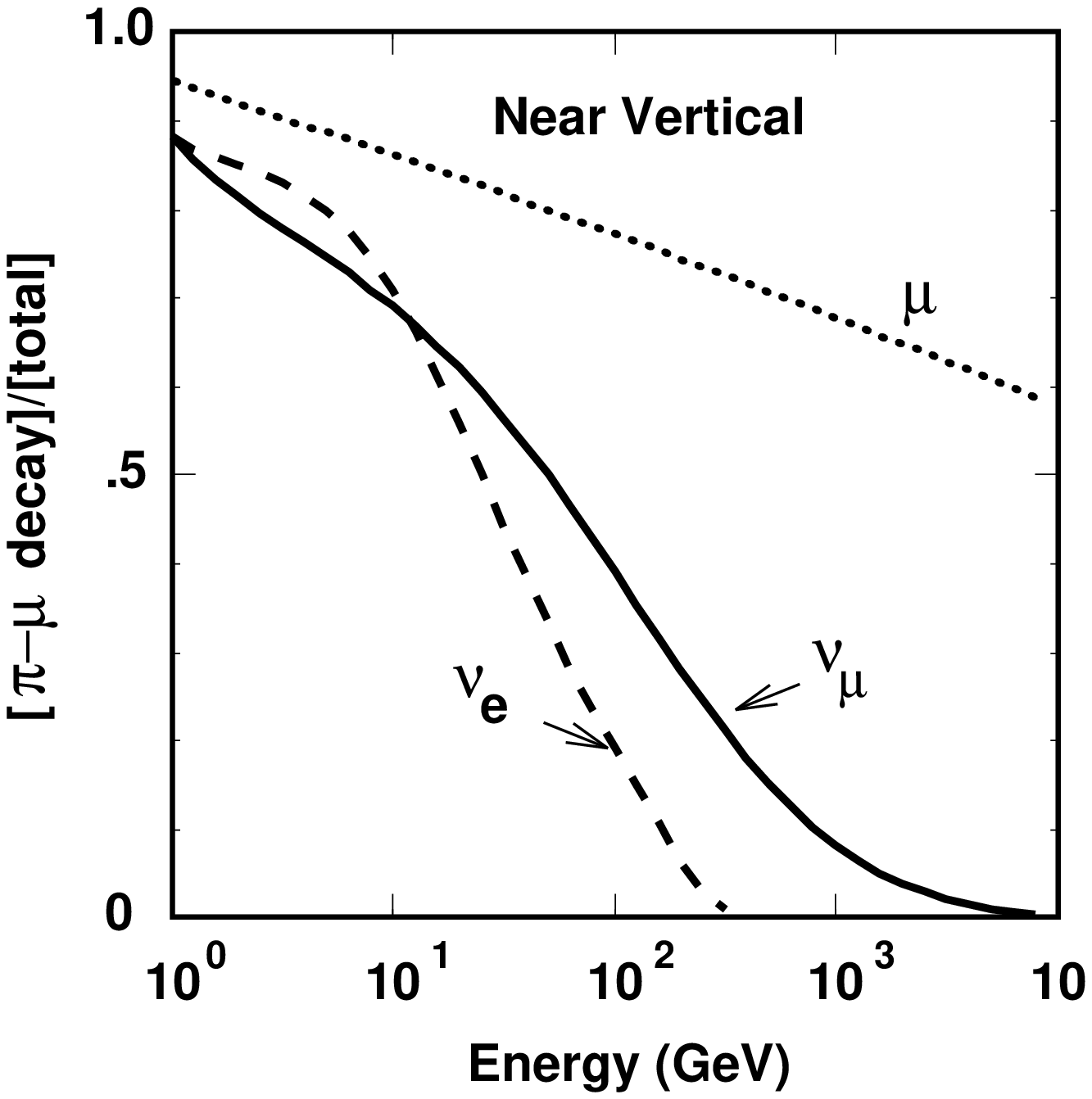,height=6.5cm}}
\caption{}
\vspace{5mm}
{\baselineskip=12pt
The ratio of $\nu_\mu + \bar\nu_\mu$,
$\nu_e + \bar\nu_e$, and $\mu^+ + \mu^-$ fluxes
produced by $\pi-\mu$--decay to the total fluxes.
Solid lines show that for $\nu_\mu + \bar\nu_\mu$,
dashed lines for $\nu_e + \bar\nu_e$,
and dotted lines for $\mu^+ + \mu^-$.
Near horizontal denotes the average over $\cos\theta = 0-0.1$ 
and near vertical denotes that over $\cos\theta = 0.9-1$.
}
\label{fig:nk-ratio}
\end{figure}

To study the portions of $\nu$'s resulting from $K$'s and 
from $\pi-\mu$-decay,
we also calculated the $\nu$--fluxes produced only with $\pi-\mu$-decay
and showed the ratio to the total flux in Fig.~\ref{fig:nk-ratio}.
The contribution of charmed mesons to atmospheric $\nu$'s 
is very small in this energy range.
The $\mu$--flux which resulted from $\pi$-decay was also calculated 
and the ratio to the total flux is also shown in the same figure. 
It is seen that the $\nu$'s created by $\pi-\mu$ decay
are the minor component above $\sim$ 30~GeV for near vertical directions
and $\sim$ 500~GeV for near horizontal directions
both for $\nu_\mu$'s and $\nu_e$'s.
It is also seen that although the main source of $\mu$--fluxes is 
$\pi$-decay up to 10,000~GeV,
the main source of atmospheric $\nu$'s is $K$'s 
above $\sim$ 30~GeV for near vertical directions and 
above $\sim$ 500~GeV for near horizontal directions.

\begin{figure}
\centerline{
\epsfile{file=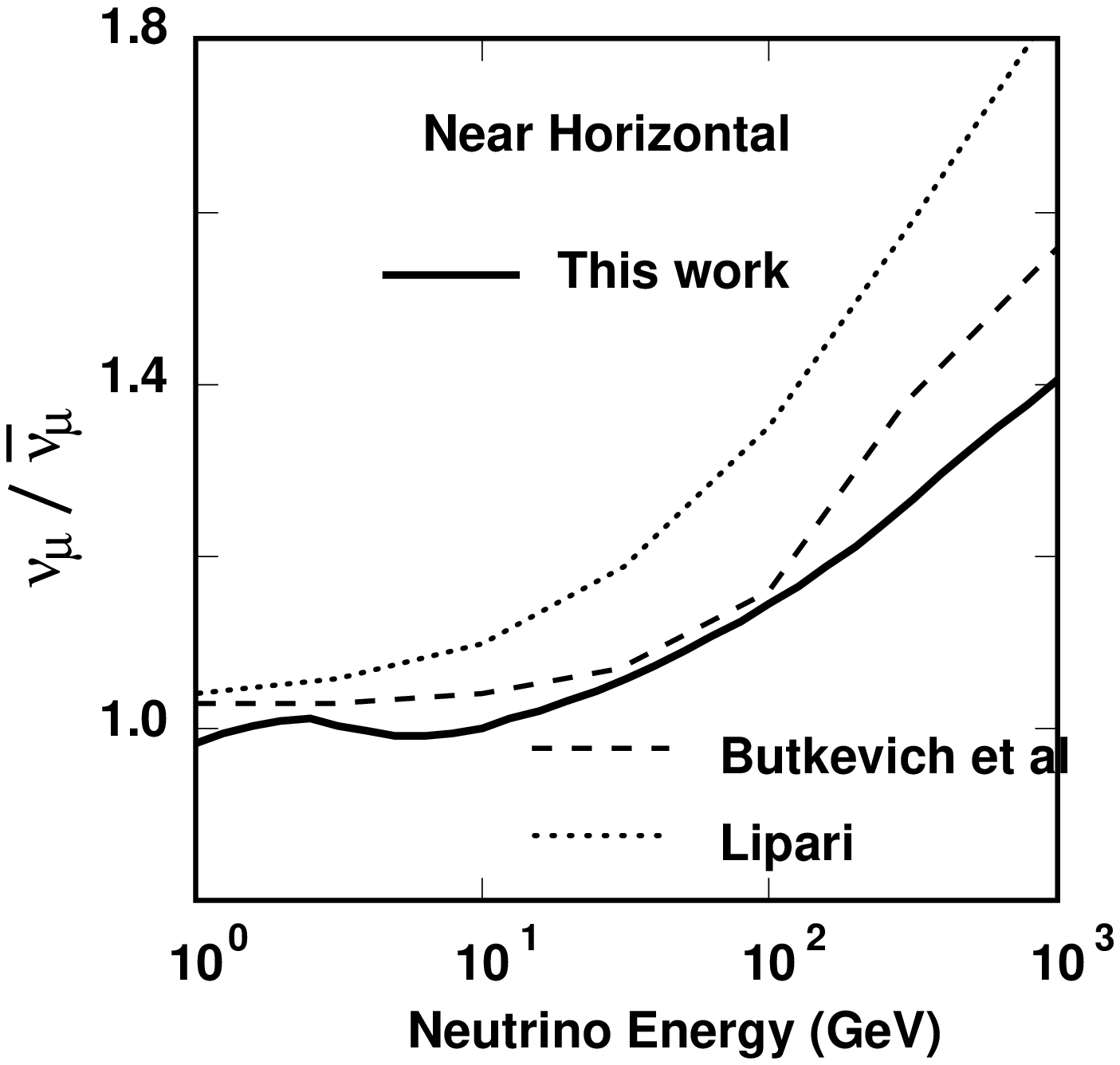,height=6.5cm}
\epsfile{file=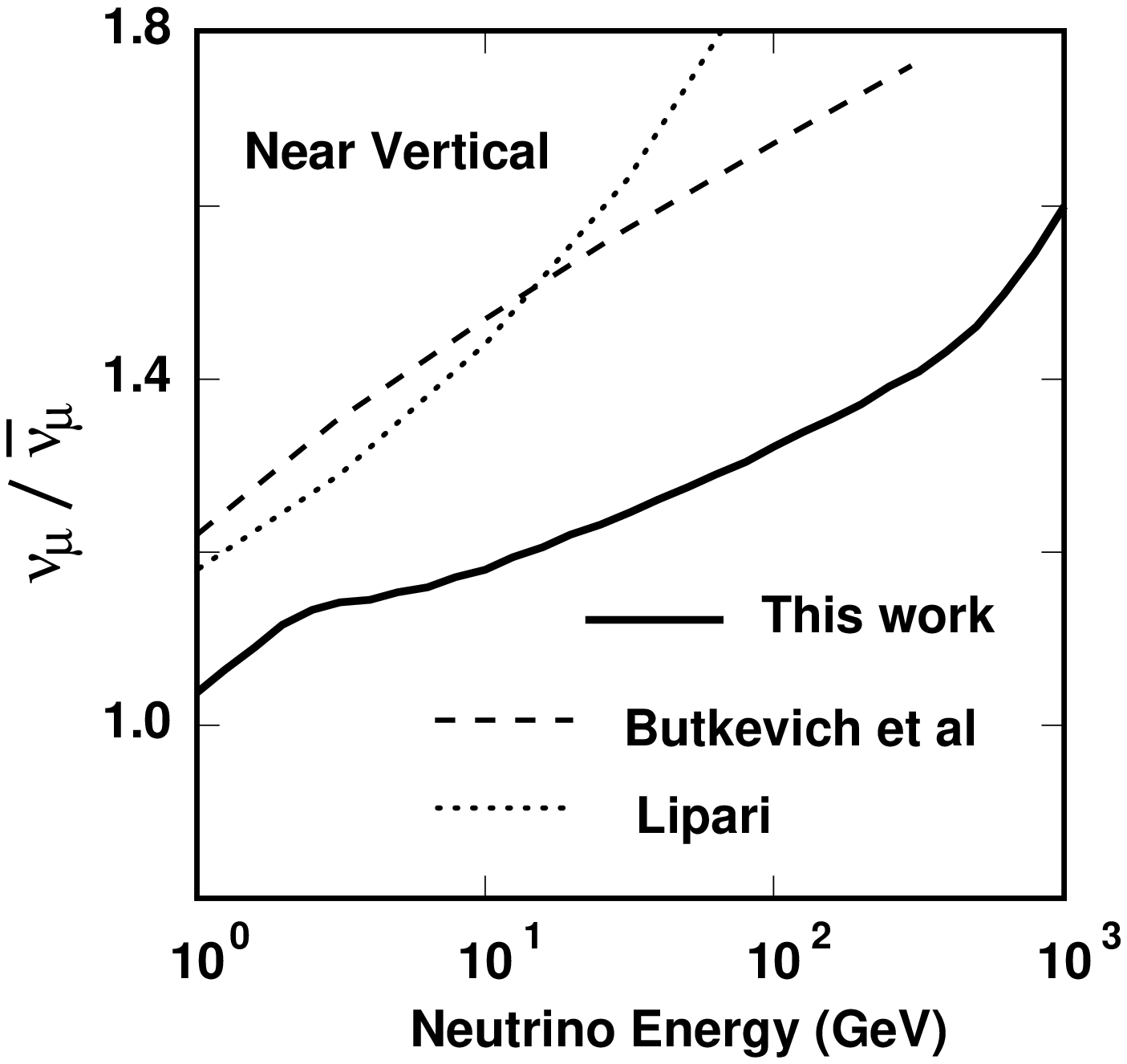,height=6.5cm}
}
\caption{}
\vspace{5mm}
{\baselineskip=12pt
$\nu_\mu /\bar \nu_\mu$--ratio.
Notations are the same as Fig.~[14].
}
\label{fig:ratio-numu-high}
\end{figure}

\begin{figure}
\centerline{
\epsfile{file=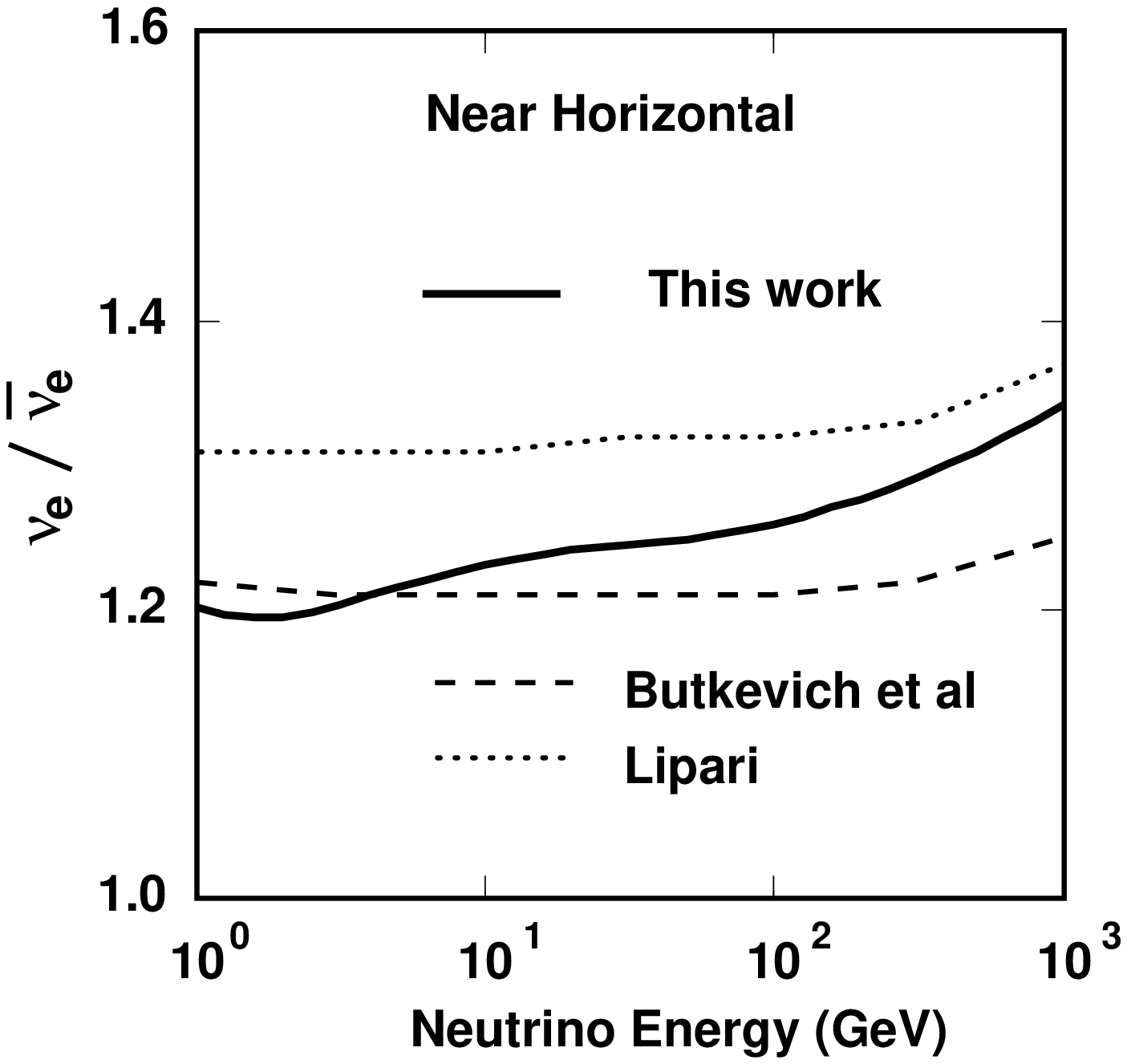,height=6.0cm}
\epsfile{file=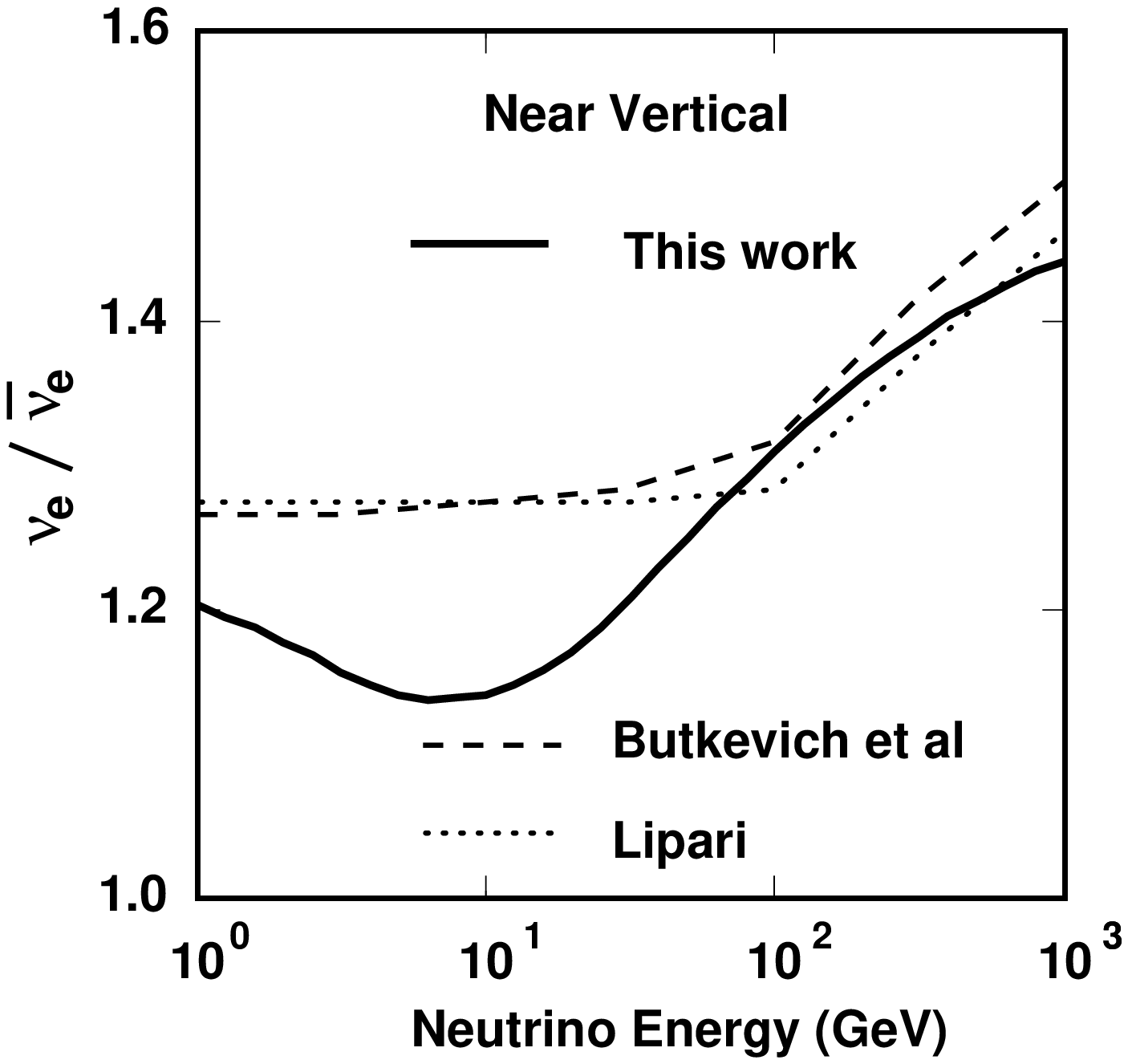,height=6.0cm}
}
\caption{}
\vspace{5mm}
{\baselineskip=12pt
$\nu_e /\bar \nu_e$--ratio.
Notations are the same with Fig.~[14].
}
\label{fig:ratio-nue-high}
\end{figure}

The ratios $\nu_\mu/\bar\nu_\mu$ and $\nu_e/\bar\nu_e$ are shown in 
in Figs~\ref{fig:ratio-numu-high} and \ref{fig:ratio-nue-high} 
for near vertical and near horizontal directions
with results of other authors.
It is difficult to determine the ratio accurately 
with the Monte Carlo method,
especially near the high energy end.
%%%%% change due to the assumption off isospin symmmetry
We do not consider the variation of $\sim 0.05$ for each ratio
to be meaningful.
Also the assumption (\ref{eq:isospin--symmetry}) is the source of 
errors in the $\nu_\mu/\bar\nu_\mu$ and $\nu_e/\bar\nu_e$ ratios
at high energies, since the main source of atmospheric 
$\nu$'s is the $K$--decay.
However, the $\nu_\mu/\bar \nu_\mu$--ratio is clearly smaller 
than other calculations~\cite{butkevich}\cite{lipari}, 
especially for near vertical directions even in the energy region
where the the assumption (\ref{eq:isospin--symmetry}) is good 
($\lesssim 30$~GeV).
For the $\nu_e/\bar\nu_e$--ratio, the agreement with other 
calculations is better than the $\nu_\mu/\bar \nu_\mu$--ratio,
but also shows significant differences at $E_\nu < 30$~GeV 
for near vertical directions.
Since we have sufficient statistics at $E_\nu < 30$~GeV
both for $\nu_\mu$ and $\nu_e$ in our Monte Carlo method,
we consider that those differences result from differences
in the calculation schemes, such as the
interaction model and/or the atmospheric density structure.
Our ratios calculated here agree well 
with those calculated in Section~\ref{sec:nflx-low} at 
$\sim 3$~GeV.

\begin{figure}
\centerline{
\epsfile{file=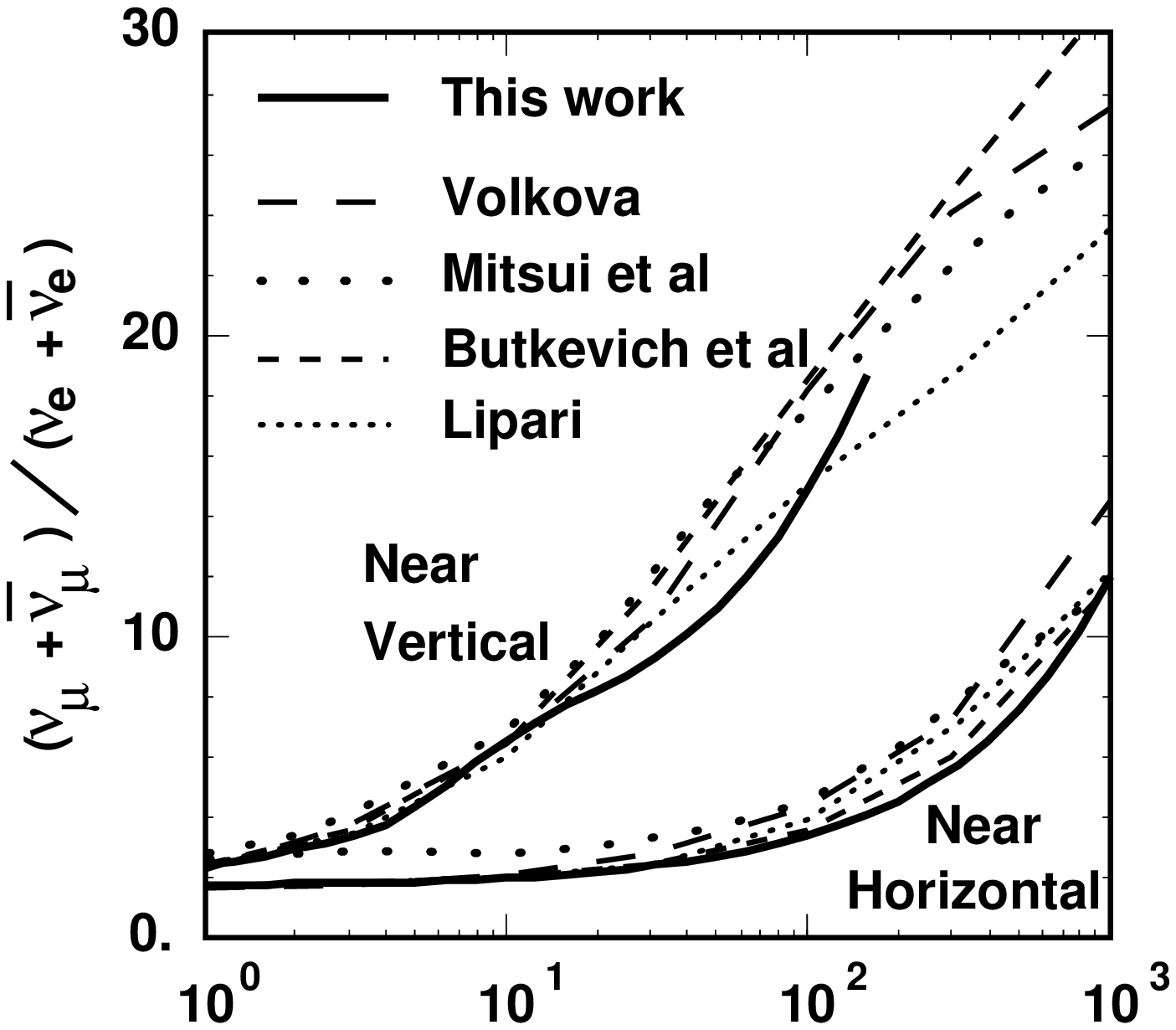,height=6.0cm}
}
\caption{}
\vspace{5mm}
{\baselineskip=12pt
$(\nu_\mu +\bar \nu_\mu)/(\nu_e +\bar \nu_e)$--ratio.
Notations are the same with Fig.~[14].
}
\label{fig:numu-nue-ratio}
\end{figure}

The ratio $(\nu_\mu +\bar \nu_\mu)/(\nu_e +\bar \nu_e)$ calculated 
here is also compared with other authors in Fig.~ \ref{fig:numu-nue-ratio}.
Our result is smaller than others for $\gtrsim 10$~GeV both
for near vertical and near horizontal directions,
reflecting the larger $\nu_e$--fluxes than other calculations
in this energy region.
This may be due to differences in the calculation scheme.
However, all results show good agreement in the $\lesssim 10$~GeV
region, except for that of Mitsui et al. for near horizontal directions.
Their result is larger than others by $\sim$ 50~\% 
in the $\lesssim 10$~GeV for near horizontal directions,
and by $10-15$~\% even for near vertical directions 
in the same energy region.
We note that Mitsui et al. did not take into account the effect
of muon polarization.
This explains the difference for near vertical directions,
but not for near horizontal directions.
There seems to be other differences in their calculation scheme.

\subsection{Flux of Atmospheric Muons}
\label{sec:mflx}

\begin{figure}
\centerline{\epsfile{file=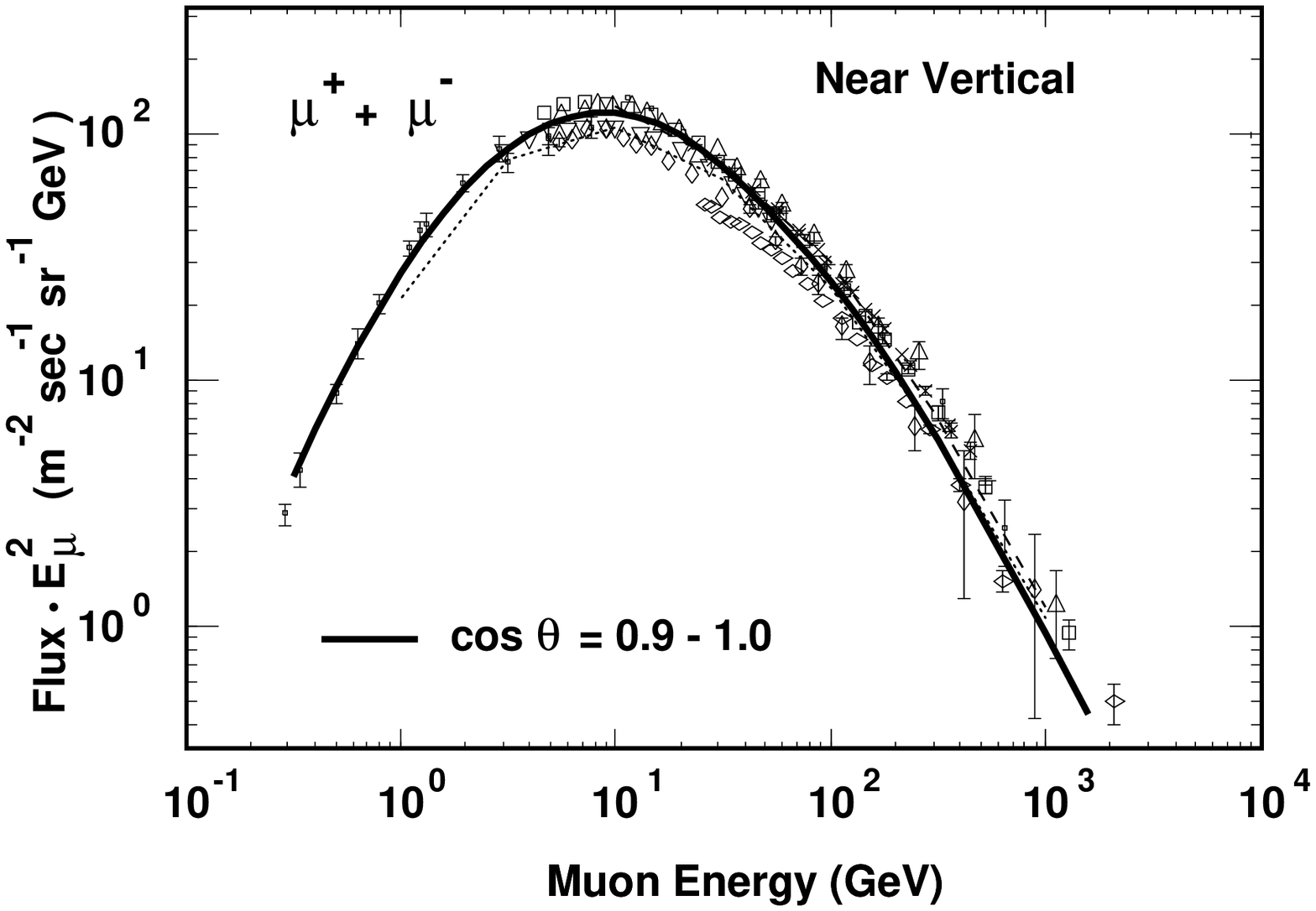,height=6.5cm}}
\caption{}
\vspace{5mm}
{\baselineskip = 12pt
Calculated $\mu^+ + \mu^-$--flux for $\cos\theta = 0.9-1$ 
and observed fluxes. 
Dots are from Ref.~[62], %[kiel]
squares from Ref.~[63], %[rastin1]
crosses from Ref.~[64], %[ayre1]
minuses from Ref.~[65], %[ayre2]
upward triangles from Ref.~[66], %[nandi1]
downward triangles from Ref.~[67], %[bateman]
vertical diamonds from Ref.~[68], %[hayman]
and
horizontal diamonds from Ref.~[69]. %[burnet]
Also shown are the calculated results from Ref.~[14] %[butkevich]
(dashed thin line) and Ref.~[15] %[lipari]
(dotted thin line) both for
$\cos\theta = 1$.
}
\label{fig:v-mu-obs}
\end{figure}

\begin{figure}
\centerline{\epsfile{file=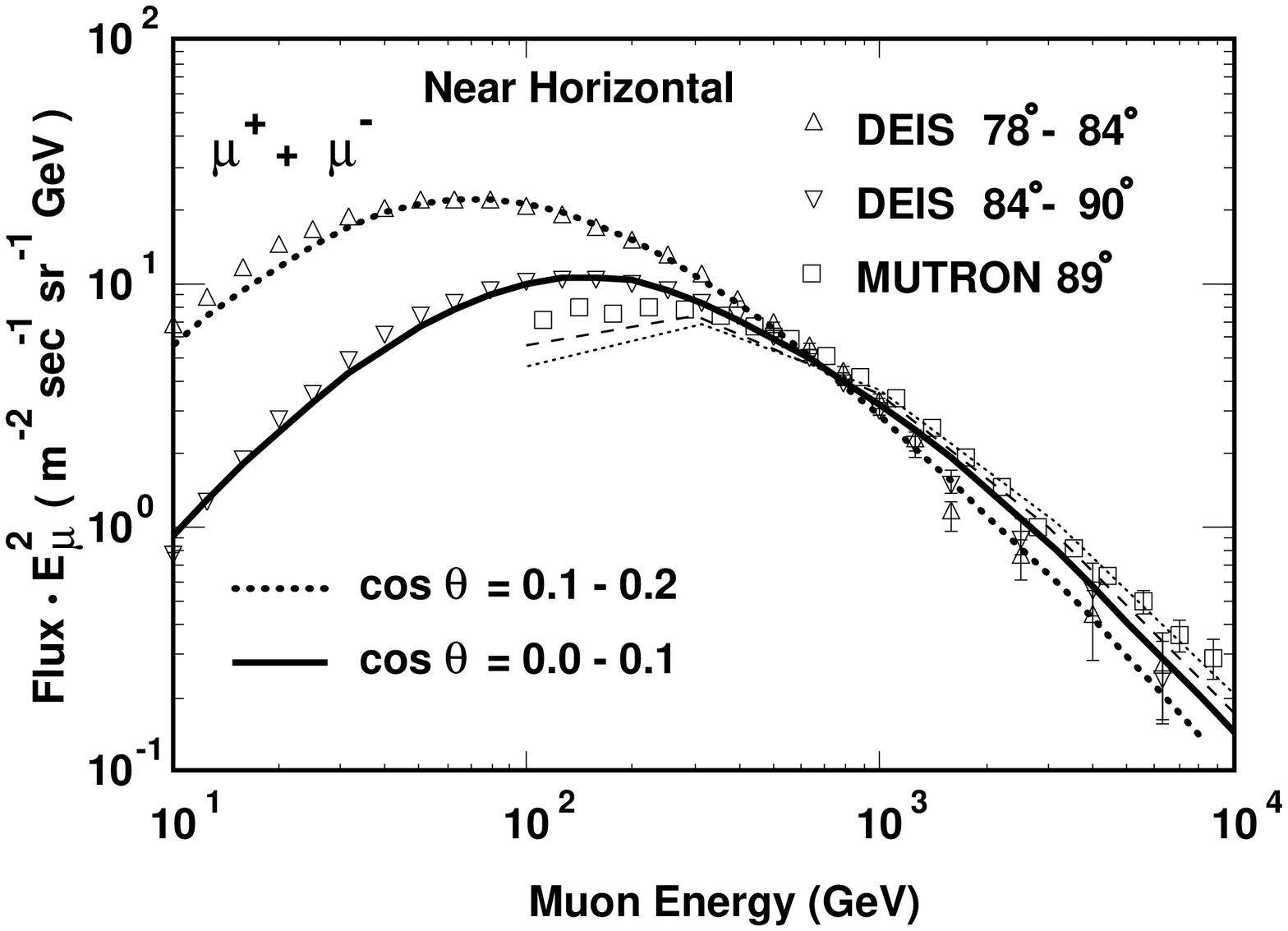,height=6.5cm}}
\caption{}
\vspace{5mm}
{\baselineskip = 12pt
Calculated $\mu^+ + \mu^-$--flux for $\cos\theta = 0-0.1$ (solid line)
and $0.1-0.2$ (dotted line),
and the observed flux by the DEIS (Ref.~[69]) 
and MUTRON (Ref.~[70]) groups.
Also shown are calculated results from Ref.~[14] %butkevich
(dashed thin line) and Ref.~[15] %lipari 
(dotted thin line) both for $\cos\theta = 0$, 
for comparison with the MUTRON data.
We note that the DEIS observation is added
for the zenith angle bins of $78^\circ - 84^\circ$ 
and $84^\circ - 90^\circ$ corresponding to 
$\cos\theta=0.1-0.2$ and $\cos\theta=0-0.1$.
}
\label{fig:h-mu-obs}
\end{figure}

There have been many observations of $\mu$--flux
for near vertical and near horizontal directions at sea level.
Since the $\mu$--flux is the complementary part of 
the $\nu$--flux in $\pi-\mu$ decay,
the calculation of $\nu$--fluxes is often 
examined by the comparison of the calculated $\mu$--flux
with the observed flux.
We calculated the atmospheric $\mu$--flux using the hybrid method;
we first calculate the $\mu$--yield function, then
integrated it with the nucleon flux.
In Fig.~\ref{fig:v-mu-obs}, we present our calculated 
$\mu^+ + \mu^-$ fluxes for near vertical directions ($\cos\theta = 0.9-1$) 
and the observed near vertical flux,
and in Fig.~\ref{fig:h-mu-obs} for near horizontal directions.
We note that although the main source is different mesons for
atmospheric $\mu$'s and $\nu$'s at high energies,
the nucleon energy dependence of the $\mu$--yield function is very 
similar to that of the $\nu_\mu$--flux above 100~GeV for all
zenith angles.

In the same figures, we show the calculated results 
of Butkevich et al.~\cite{butkevich} 
and Lipari~\cite{lipari}.
We used the result of $\cos\theta = 1$ 
for the near vertical direction and of $\cos\theta = 0$ 
for near horizontal directions
for these authors.
Since the DEIS group measured the  $\mu$--flux
at many zenith angles,
we summed their data in the zenith angle bins of 
$78^\circ - 84^\circ$ 
and  $84^\circ - 90^\circ$ corresponding to 
$\cos\theta=0.1-0.2$ and $\cos\theta=0-0.1$.
The MUTRON group measured the muon flux in the direction of 
$86^\circ - 90^\circ$
and the flux averaged zenith angle is $89^\circ$.

It can be seen that the agreement of our calculation and the
observations is good, although there is some variation in
observed fluxes among the different groups.
Our calculation agrees with DEIS data within $\sim$~15~\% 
at all energies, and $\lesssim$~5~\% for 100 -- 1,000~GeV.
The agreement is especially good for the near horizontal directions.
Taking into account the fact that average zenith angle of 
the MUTRON data is $89^\circ$,
the agreement of MUTRON data and our calculation is also 
very good.
However, since the main parent mesons are different for $\nu$'s 
and $\mu$'s,
these agreements shown above do not fully justify
our calculation of $\nu$--fluxes.

\begin{figure}
\centerline{
\epsfile{file=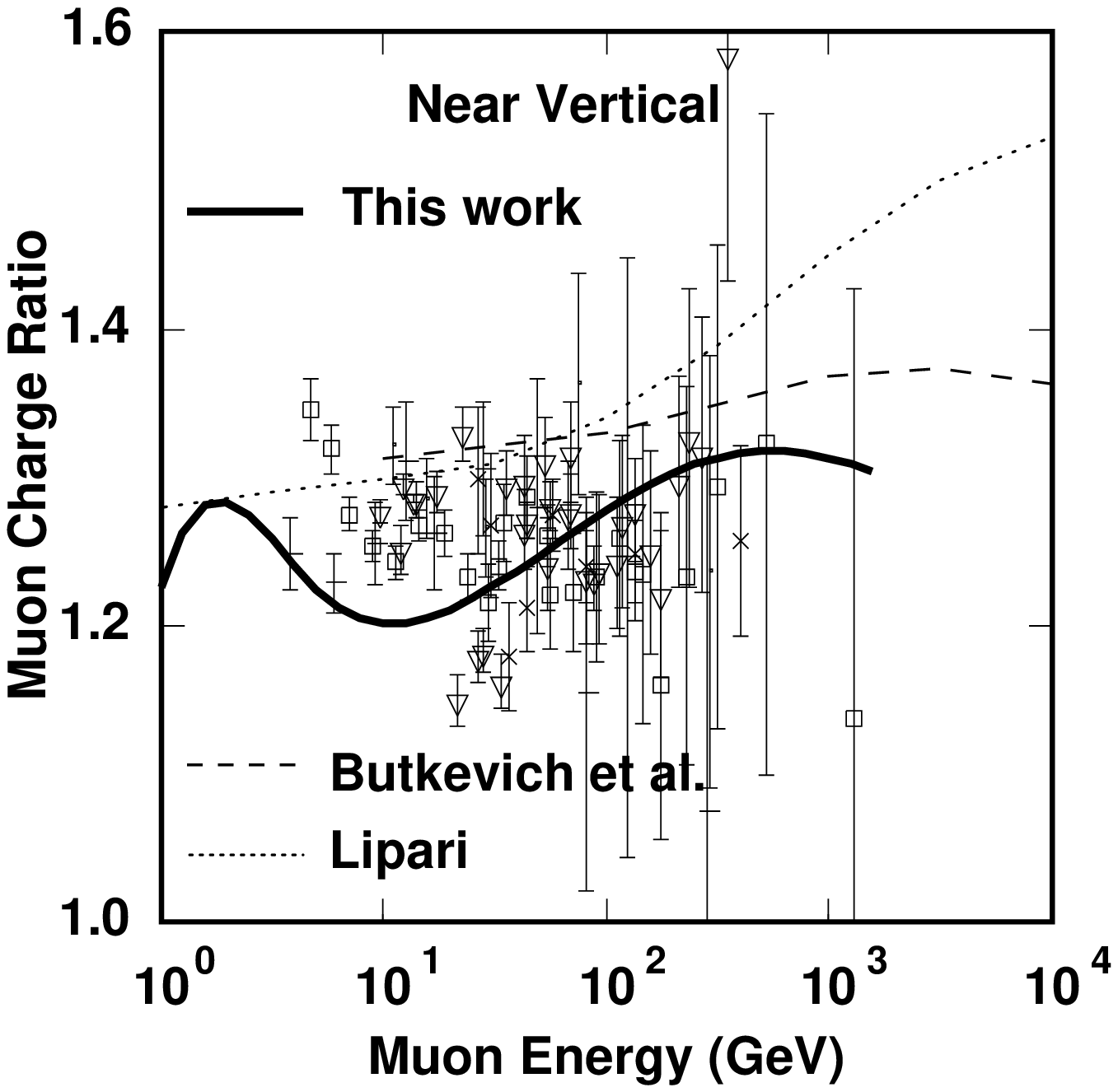,height=6.5cm}
\epsfile{file=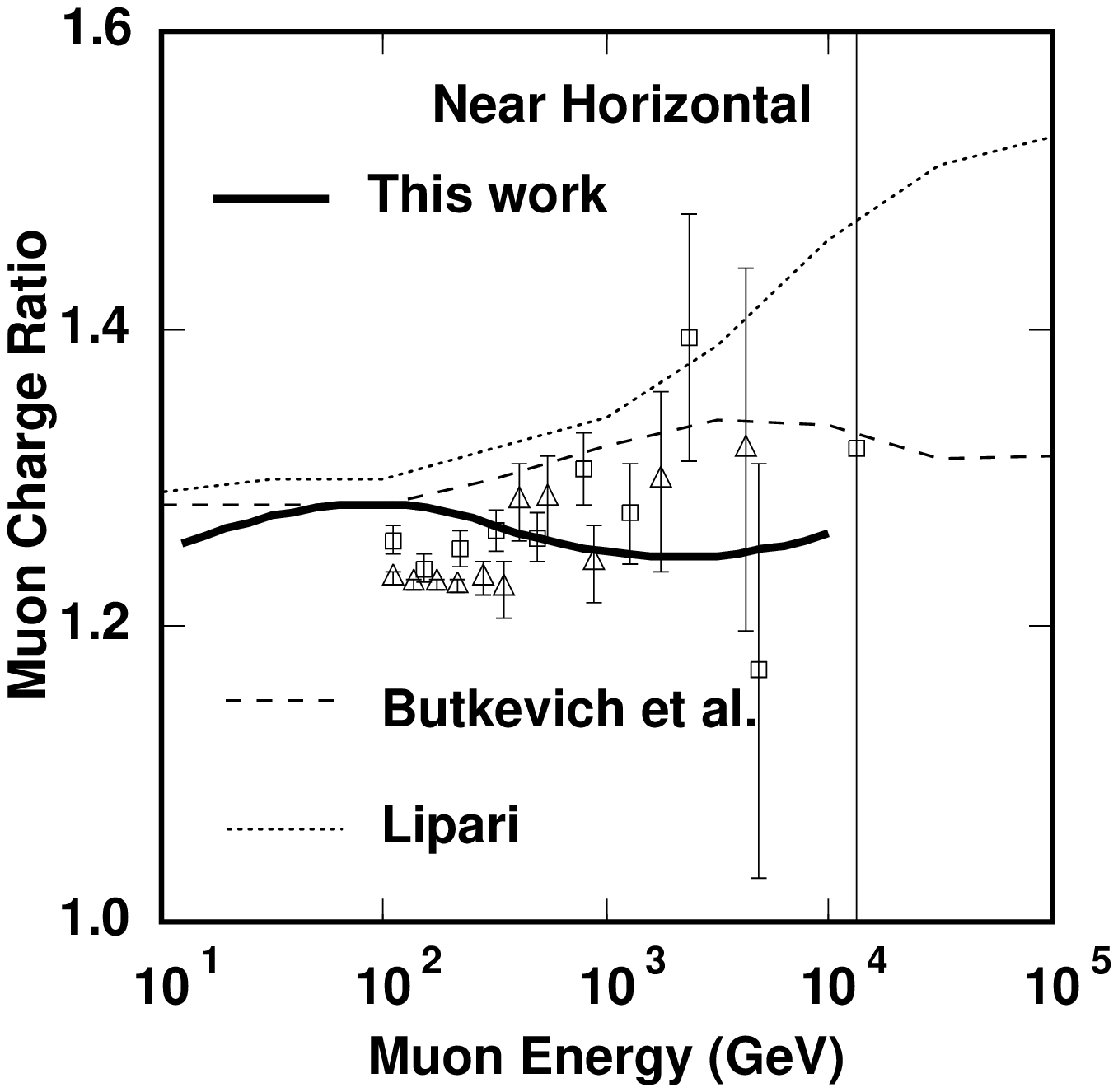,height=6.5cm}
}
\caption{}
\vspace{5mm}
{\baselineskip = 12pt
Comparison of calculated 
$\mu^+/\mu^-$--ratios with observations.
For the near vertical directions,
dots are from Ref.~[62], %[kiel]
crosses from Ref.~[69], %[burnett]
squares from Ref.~[72], %[rastin2]
minuses from Ref.~[73], %[appleton]
upward triangles from Ref.~[74], %[nandi2]
and diamonds from Ref.~[75]. %[baxen]
For the near horizontal directions,
triangles are Ref.~[70] %[deis]
and squares from Ref.~[71]. %[matsuno]
Also for near horizontal direction,
calculated $\mu^+/\mu^-$--ratios are averaged for 
$\cos\theta =0-0.2$ for the results of Lipari and this work.
}
\label{fig:mu-ratio}
\end{figure}

%==================
The charge ratio $\mu^+/\mu^-$ was also calculated and shown 
in Fig.~\ref{fig:mu-ratio} with the result of other 
authors~\cite{butkevich}\cite{lipari}. 
As we do not consider that the variation of $\lesssim 0.05$
to be meaningful in our calculation,
a constant value of $\sim 1.25$ can explain our results 
both for near vertical and near horizontal directions.
Other calculations show an increase with energy, especially 
for the near vertical directions.
%%%%
We note that the assumption (\ref{eq:isospin--symmetry}) 
is almost valid for $\mu^+/\mu^-$--ratio, since
the main source is still the $\pi-\mu$--decay,
and the experimental results do not show 
such an increase but are consistent with the constant value 
$\sim 1.25$ for both directions.

\begin{figure}
\centerline{\epsfile{file=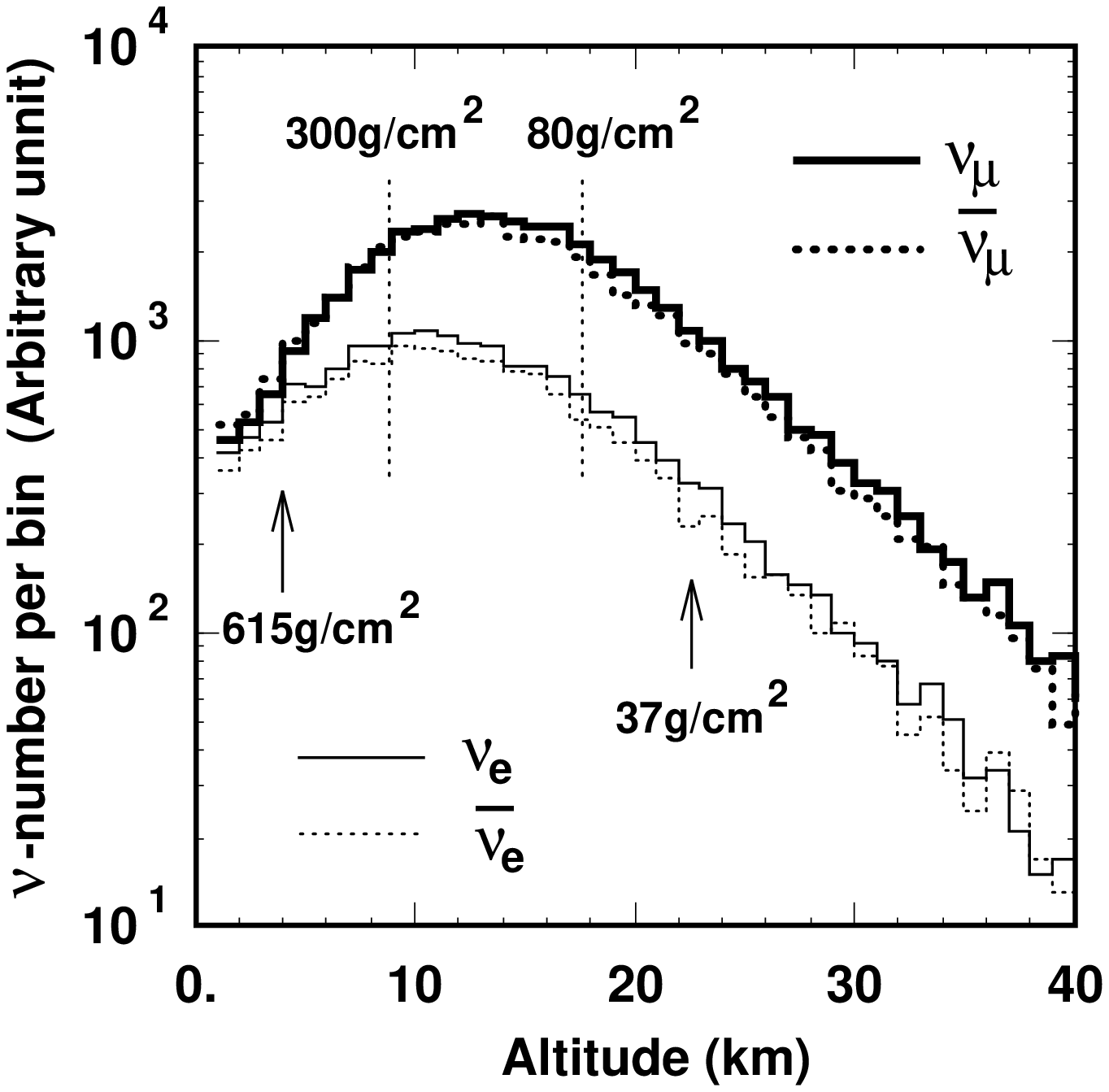,height=6.5cm}}
\caption{}
\vspace{5mm}
{\baselineskip=12pt
Production height distribution of $\nu$'s 
with energy $\ge 400$~MeV
by vertical cosmic rays.
}
\label{fig:height-distribution}
\end{figure}

At low energies,
the production altitude of $\nu$'s is 80 -- 300 ${\rm g/cm^2}$ 
for vertical directions, which corresponds 
to 9 -- 18~km 
(Fig.~\ref{fig:height-distribution}) altitude.
Since $\mu$'s travel $\sim$~6~km before decay at 1~GeV on average,
$\mu$'s observed at sea level are not directly related to 
$\nu$'s at low energies ($\lesssim 1$~GeV). 
For the examination of the calculated atmospheric $\nu$--flux 
at low energies,
the observation of $\mu$--fluxes at the production height 
is necessary.

\begin{figure}
\centerline{\epsfile{file=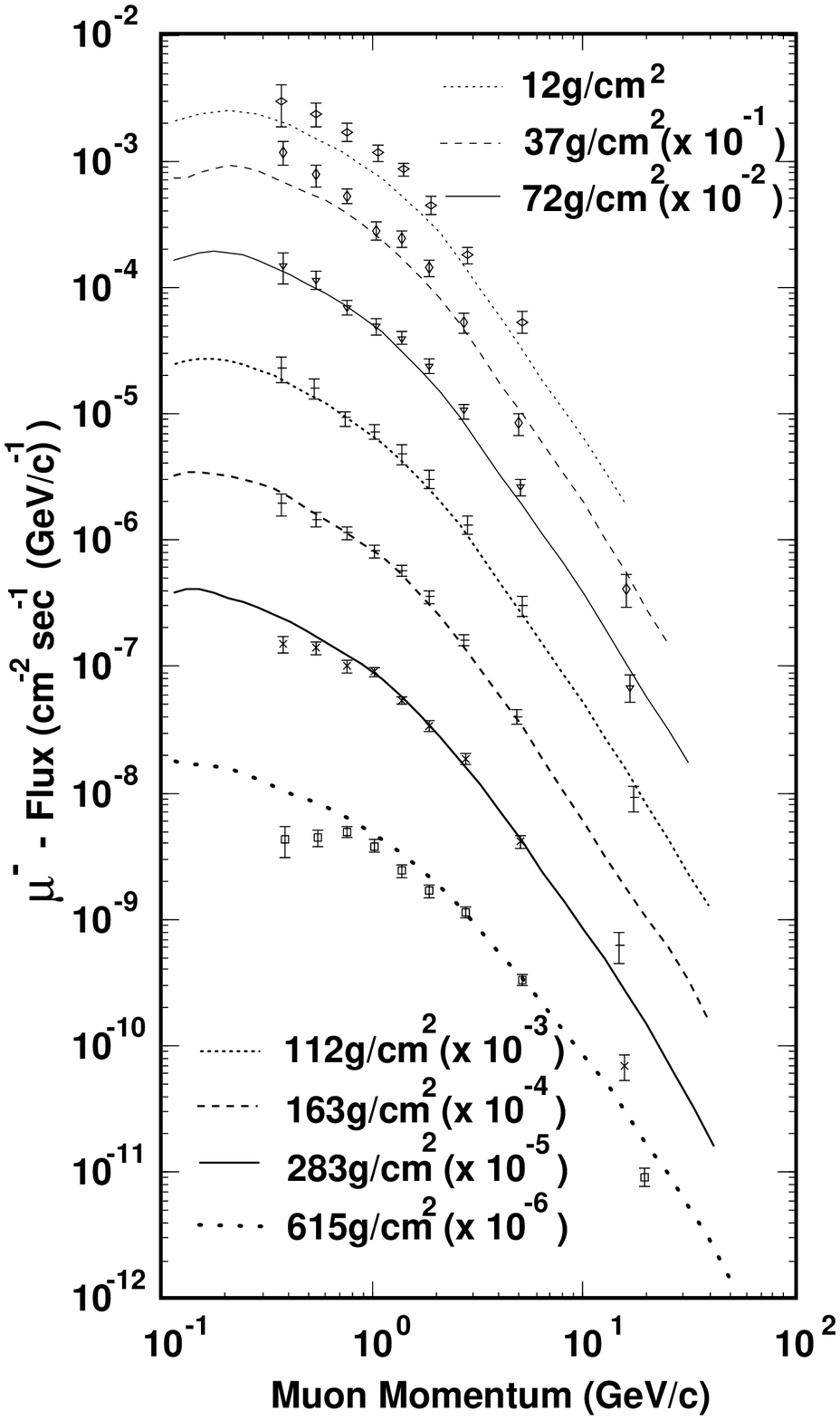,height=12cm}}
\caption{}
\vspace{5mm}
{\baselineskip=12pt
A comparison of $\mu^-$--fluxes observed 
by the MASS (Matter-Antimatter Superconducting Spectrometer) 
experiment~[76] %[marco]
at several altitudes and this work.
}
\label{fig:marco}
\end{figure}

Recently, the $\mu^-$--flux has been measured 
with good accuracy by the MASS (Matter-Antimatter 
Superconducting Spectrometer) experiment~\cite{marco} at high 
altitudes.
In Fig.~\ref{fig:marco}, we compared the observed $\mu^-$--fluxes 
and our calculation at the same altitudes.
This $\mu$--flux was calculated by the full Monte Carlo method, 
the same as the low energy $\nu$--flux.
The agreement of experiment and calculation is very good
except for very high altitudes ($ \lesssim 37$~$g/cm^2$)
and the low momentum region at low altitudes (615~$g/cm^2$).
We note that since the low energy atmospheric $\nu$'s are 
created at the altitude of 80 -- 300~$g/cm^2$,
the contribution of atmospheric $\nu$'s created at
very high altitude ($\lesssim 37$~$g/cm^2$)
or low altitude (615~$g/cm^2$) to the total 
$\nu$--flux is relatively small. 
(See Fig.~\ref{fig:height-distribution}.)

\section{The Systematic Error and Other Uncertainties}
\label{sec:syserror}

The systematic error in the atmospheric $\nu$--fluxes
comes mainly from the uncertainty of the cosmic ray primary flux.
Even at low energies, 
where the primary cosmic ray flux is rather well studied, 
it is difficult to determine the absolute value due to
the uncertainties in the instrumental 
efficiency ($\sim$ 12\%) and exposure factor (2 -- 3\%)~\cite{seo1}.
These uncertainties in the primary cosmic ray flux increase with 
energy.
In our compilation,
the error in the fit is $\sim$ 10~\% for the 
nucleon flux at 100~GeV and $\sim$~20~\% at 100~TeV.
Assuming $\sim$~10\% uncertainty below 100~GeV,
the systematic error in the atmospheric $\nu$--fluxes is 
estimated to be $\sim$ 10\% at $\le 3$~GeV, 
increasing to $\sim$~20~\% at 100~GeV, 
and remain almost constant up to 1,000~GeV.

We note that the uncertainty of the primary cosmic ray flux 
increases more rapidly than the fitting error 
above 10~TeV$/$nucleon.
The JACEE group suggests a steepening of the cosmic ray proton 
spectrum above 40~TeV~\cite{jacee-p}.
Using this steep proton spectrum, 
the atmospheric $\nu$--flux decreases 3--4~\% at 1,000~GeV, 
$\sim$ 10~\% at 3,000~GeV, and $\gtrsim 20$~\% above 
10,000~GeV.
Below 300~GeV, the difference is negligibly small.

The uncertainty in the nucleon spectrum above 100~TeV, where
almost no direct observations are available, is crucial
for the calculation of atmospheric $\nu$'s above 1,000~GeV,
The air shower technique, which is commonly used to 
study cosmic rays above 100~TeV,
can not determine the chemical composition with the accuracy 
we need.
The nucleon spectrum could be very different from that assumed here, 
depending on the major chemical component of cosmic rays above 100~TeV.

The interaction model is another source of systematic errors.
In our comparison, the agreement of the LUND model and the COSMOS code
with the experimental data is $\lesssim$~10~\%.
The agreement of the NUCRIN code and experimental data is not as good 
as the LUND code.
The authors of the NUCRIN code claim that the agreement is 
within 10 -- 20~\%~\cite{nuc2}.
However, the hadronic interaction below 5~GeV is related to very
low energy atmospheric $\nu$--fluxes;
when we switch off the hadronic interaction for $\le$ 5~GeV,
they decrease 2 -- 5~\% at 1~GeV, 15 -- 25~\% at 300~MeV, and 
45 -- 55~\% at 100~MeV dependeng on the rigidity cutoff.
Therefore, we may conclude that the error of the NUCRIN code does
not affect the atmospheric $\nu$--flux by more than 10~\%
in our calculation for $\ge$ 300~MeV.
The systematic error caused by the hadronic interaction model is
estimated to be $\sim$ 10~\% above 300~MeV.

The calculation scheme -- the one-dimensional approximation in all 
energy regions and the superposition model for nucleus--nucleus
interactions at high energies -- is also 
the possible source of some systematic error.
Since one--dimensional approximation is justified at high 
energies, it is expected to be accurate above 3~GeV.
With the one--dimensional approximation, however,
the calculation of rigidity cutoff is very simplified,
and this may result in a systematic error in the absolute value of the
atmospheric $\nu$--fluxes of 10 -- 20~\% at 100~MeV
and 5~\% at 1~GeV.
Other effects caused by the one--dimensional approximation 
and errors due to the superposition model at high energies are
considered to be small compared to other errors.

%%% uncetainty with monte carlo calculation
The calculation method also results in an error in 
the atmospheric $\nu$--flux.
With the full Monte Carlo calculation, 
it is estimated to be $\lesssim 2 - 3$~\% for 30~MeV -- 3~GeV 
due to the statistics.
With the hybrid method, the statistics and the fitting 
error are sources of the uncertainty.
Both the statistics and fitting error are combined and estimated to be 
$\lesssim 5$~\% up to 100 -- 300~GeV for $\nu_\mu$ and $\bar\nu_\mu$, 
and up to 30 -- 100~GeV for $\nu_e$ and $\bar\nu_e$,
depending on the zenith angle.

Combining all the systematic and non--systematic errors,
the total error is estimated as 
20~\% at 100~MeV, 15 \% from 1~GeV to 100~GeV, and 20 -- 25~\% at 
the highest energy in our calculation.
However, 
the errors of the species ratio is smaller than the absolute value,
since the $\nu$--species ratio is not affected much 
by the uncertainty of primary fluxes and the calculation scheme. 
It is estimated to be
$\lesssim$ 10~\% below 100~GeV for $\nu/\bar\nu$ and
$\lesssim$ 5~\% below 30~GeV for $(\nu_\mu + \bar\nu_\mu)/(\nu_e + \bar\nu_e)$.
These errors also increase to 10 -- 15~\% at the highest energies
in our calculation.

Although the main parent meson is different for 
high energy atmospheric $\mu$'s and $\nu$'s,
one may consider that
the comparison of calculated and observed
atmospheric $\mu$--flux reduces the systematic error 
due to the primary cosmic ray flux.
The agreement of our calculation,  DEIS data,
and MUTRON data for near horizontal directions (Fig.~\ref{fig:h-mu-obs}) 
suggests that the systematic error of the atmospheric 
$\nu$--flux caused by the uncertainty of the cosmic ray flux may be 
$\lesssim$~10~\% at $E_\nu =$ 100 -- 1,000~GeV in our calculation.
However, we note that
there are similar problems for observation of $\mu$--fluxes
with that of primary cosmic rays: determining
the instrumental efficiency and the exposure factor.
The MASS group claims that their total uncertainty in
efficiency is around 20~\%.
This systematic error for ground based experiments
could be smaller, but
it is seen in Fig.~\ref{fig:v-mu-obs} that
$\mu$-fluxes observed for near vertical directions
by different groups differ by more than 20~\%.
Also taking the possible systematic error in the $K/\pi$-ratio 
into account,
we conclude that the systematic error caused by 
the uncertainty in the  primary flux is $\sim$ 20~\%
at $E_\nu = 1,000$~GeV.

\section{Summary}
\label{sec:summary}

In this paper, we presented the calculation of 
atmospheric $\nu$--fluxes as follows;
first, we summarized the physics related to the primary cosmic ray flux.
The low energy cosmic ray flux was parametrized following the
work of Nagashima et al.~\cite{nagashima} based on the compilation 
of the cosmic ray spectrum by Webber and Lezniak~\cite{wl1} 
for solar max., mid., and min.
In order to calculate the rigidity cutoff due to the 
geomagnetic field,
we simulated the trajectories of cosmic rays.
Also the primary cosmic ray flux in the 100~GeV -- 100~TeV range
was compiled for H, He, CNO, Ne-S, Fe nuclei groups 
calculating the nucleon flux for protons and neutrons.
We used the NUCRIN interaction model~\cite{nuc1}\cite{nuc2} 
for $<$ 5~GeV,
LUND model (JETSET ver. 6.3)~\cite{lund1}\cite{lund2} 
for 5 -- 500~GeV, and
COSMOS for $>$ 500~GeV for the hadronic interaction 
interactions of cosmic rays.
The atmospheric $\nu$--flux was calculated with a
full Monte Carlo method for 30~MeV -- 3~GeV,
and with a hybrid method for 1~GeV -- 1,000~GeV.

One of the most important results for low-energy $\nu$--fluxes 
is that the ratio $(\nu_e + \bar \nu_e)/(\nu_\mu + \bar \nu_\mu)$
is almost the same ($\sim 0.5$) as other calculations for 
100~MeV -- 3~GeV~\cite{bar89}\cite{bn89}\cite{lk90}\cite{perkins},
whereas underground detectors found a
significant difference in the ratio 
(e--like~event)$/$($\mu$--like~event) 
from the value expected from the calculated atmospheric
$\nu$--flux.
We note that 
the quantity $(\nu_e + \bar \nu_e)/(\nu_\mu + \bar \nu_\mu)$
remains relatively unaffected by variations in the
interaction model and primary cosmic ray spectrum including 
the chemical composition.
The difference between the observed and the expected value of
the ratio (e--like~event)$/$($\mu$--like~event) 
might be explained by other physics, 
such as $\nu$--oscillations.

If this difference is to be explained by
$\nu$--oscillations with $\Delta m^2 \sim 10^{-2}~{\rm eV}^2$,
the up-going $\mu$-flux, 
which is induced by the high energy $\nu$-flux,
will show a different zenith angle dependence
from the expected one.
We calculated 
the zenith angle dependence of the atmospheric $\nu$--flux in 
detail: for each zenith angle bin of 
$\cos\theta = 0 - 0.1, 0.1 - 0.2, \cdots 0.9 -1.0$,
for 1 -- 1,000~GeV.
This atmospheric $\nu$--flux could be used to calculate 
the expectation flux of up-going $\mu$'s.
We note that the absolute value and the ratio are connected 
to the lower values smoothly at $\sim$ 3~GeV.
Also they are compared with the calculation of other 
authors~\cite{volkova}\cite{mitsui}\cite{butkevich}
\cite{lipari}.

Atmospheric $\mu$--fluxes were also calculated at sea level 
and at high altitudes.
They were compared with the experimental data
and the agreements are found to be satisfactory.
The agreement at high altitude is especially important for 
the calculation of the atmospheric $\nu$--flux at low energies.
Although the parent mesons are different for atmospheric $\nu$'s
and $\mu$'s, 
our calculation, DEIS data, and MUTRON data agree very well
each other for the $\mu$--flux of near horizontal directions 
at high energies ($\gtrsim 100$~GeV).
We conclude that we have used a reasonable primary cosmic ray 
spectrum, chemical composition, and interaction model.

We stress again that the main source of the systematic error in
the atmospheric $\nu$--flux is the uncertainty of the
primary cosmic ray flux. 
Especially for the calculation of the atmospheric 
$\nu$--flux above 1,000~GeV, 
the lack of knowledge of the cosmic ray flux above 
100~TeV$/$nucleon is crucial.

\section{Acknowledgments}
We are grateful to P.G.~Edwards for his careful reading of
the manuscript and comments.
The numerical calculation were performed the FACOM M780
in the computer room of the Institute for Nuclear Study,
University of Tokyo.

\newpage

% tables follow here

\begin{table}
\caption{Fitted parameters for each chemical component of cosmic rays.}
\label{table:fit-parm}
\begin{tabular}{c c c}
Nucleus & $A~({\rm m^{-2} sec^{-1} sr^{-1} GeV^{-1}})$ & $\gamma$\\ \hline
H & $6.65 \pm 0.13 \times 10^{-2}$ & $-2.75 \pm 0.020$\\
He &$3.28 \pm 0.05 \times 10^{-3}$ & $-2.64 \pm 0.014$\\
CNO &$1.40 \pm 0.07 \times 10^{-4}$ & $-2.50 \pm 0.06$\\
Ne-S &$3.91 \pm 0.03 \times 10^{-5}$ & $-2.49 \pm 0.04$\\
Fe &$1.27 \pm 0.11 \times 10^{-5}$ & $-2.56 \pm 0.04$\\
\end{tabular}
\end{table}

\vspace{1pc}

\begin{table}
\caption{The low energy $\nu$--flux for Kamioka 
(solar mid., ${\rm m^{-2} sec^{-1} sr^{-1} GeV^{-1}}$)}
\label{low-nu-kam}
\begin{tabular}{c r r r r}
 $E_\nu({\rm GeV})$& $\nu_\mu$ & $\bar \nu_\mu$ & $\nu_e$ & $\bar \nu_e$\\
\hline
 3.162$\times 10^{-2}$ &13845 &14518  &8616  &8345\\
 3.981$\times 10^{-2}$ &14645 &15097  &8493  &8232\\
 5.012$\times 10^{-2}$ &13805 &13856  &7238  &6853\\
 6.310$\times 10^{-2}$ &11080 &11154  &5597  &4995\\
 7.943$\times 10^{-2}$ & 9113 & 9143  &4539  &3940\\
 1.000$\times 10^{-1}$ & 7603 & 7568  &3767  &3272\\
 1.259$\times 10^{-1}$ & 5913 & 5903  &2916  &2495\\
 1.585$\times 10^{-1}$ & 4372 & 4372  &2192  &1851\\
 1.995$\times 10^{-1}$ & 3124 & 3120  &1597  &1341\\
 2.512$\times 10^{-1}$ & 2188 & 2180  &1105  & 933\\
 3.162$\times 10^{-1}$ & 1494 & 1486  & 752  & 636\\
 3.981$\times 10^{-1}$ &  994 &  988  & 501  & 424\\
 5.012$\times 10^{-1}$ &  647 &  641  & 322  & 274\\
 6.310$\times 10^{-1}$ &  412 &  407  & 204  & 173\\
 7.943$\times 10^{-1}$ &  256 &  252  & 127  & 107\\
 1.000    &    155 &    152  &  75.7  &  63.9\\
 1.259    &   90.9 &   87.9  &  44.3  &  37.4\\
 1.585    &   52.3 &   49.7  &  25.2  &  21.0\\
 1.995    &   29.6 &   27.9  &  13.7  &  11.1\\
 2.512    &   16.4 &   15.4  &   7.10  &   5.96\\
 3.162    &    8.92 &   8.32  &  3.65  &   3.19\\
\end{tabular}
\end{table}
\vspace{1pc}

\begin{table}
\caption{The low energy $\nu$--flux for IMB 
(solar mid., ${\rm m^{-2} sec^{-1} sr^{-1} GeV^{-1}}$)
}
\label{low-nu-imb}
\begin{tabular}{c r r r r}
 $E_\nu({\rm GeV}) $& $\nu_\mu$ & $\bar \nu_\mu$ & $\nu_e$ & $\bar \nu_e$\\
\hline
 3.162$\times 10^{-2}$ &26050& 26869&15971&14170\\
 3.981$\times 10^{-2}$ &27302& 27840&15725&13984\\
 5.012$\times 10^{-2}$ &25317& 25453&13594&11816\\
 6.310$\times 10^{-2}$ &20437& 20650&10778& 8623\\
 7.943$\times 10^{-2}$ &16790& 16744& 8758& 6765\\
 1.000$\times 10^{-1}$ &13807& 13543& 7129& 5545\\
 1.259$\times 10^{-1}$ &10535& 10430& 5421& 4155\\
 1.585$\times 10^{-1}$ & 7509&  7504& 3955& 3019\\
 1.995$\times 10^{-1}$ & 5140&  5153& 2779& 2135\\
 2.512$\times 10^{-1}$ & 3494&  3495& 1868& 1441\\
 3.162$\times 10^{-1}$ & 2243&  2260& 1207&  932\\
 3.981$\times 10^{-1}$ & 1389&  1404&  758 &  590\\
 5.012$\times 10^{-1}$ &  875&   861&  469 &  375\\
 6.310$\times 10^{-1}$ &  528&   524&  280 &  223\\
 7.943$\times 10^{-1}$ &  309&   312&  163 &  129\\
 1.000    &  183&   178&   95.0 &   77.5\\
 1.259    &  103&   100&   52.3 &   42.3\\
 1.585    &   56.6&  55.9& 28.1 &   22.5\\
 1.995    &   31.8&  30.5& 15.5 &   13.5\\
 2.512    &   17.3&  16.1&  7.86 &    6.64\\
 3.162    &    9.22&  8.43 &3.86 &    3.06\\
\end{tabular}
\end{table}
\vspace{1pc}

\begin{table}
\caption{$\nu_\mu-$flux $\times E_\nu^3$ 
(${\rm m^{-2} sec^{-1} sr^{-1} GeV^2}$) 
calculated with the hybrid method.
The value above $1\times 10^4$ is the smooth extension
with a power fit.
}
\label{s-numu3}
\begin{tabular}{c c c c c c c c c c c}
% s-numu.d3
 $E_\nu({\rm GeV})\backslash \cos\theta$&$ .0-.1$&$ .1-.2$&$ .2-.3$&$ .3-.4$&$ .4-.5$&$ .5-.6$&$ .6-.7$&$ .7-.8$&$ .8-.9$&$ .9-1.$\\
\hline
1.000&213.5&208.0&202.1&195.8&189.1&182.2&176.1&170.7&165.9&161.9\\
1.259&245.1&237.3&229.2&221.0&212.6&204.3&196.9&190.5&185.0&180.4\\
1.585&275.5&264.5&253.8&243.4&233.4&223.8&215.4&208.2&202.1&197.0\\
1.995&302.8&288.0&274.2&261.5&249.8&239.0&229.7&221.9&215.4&210.2\\
2.512&327.1&308.3&291.5&276.4&262.8&250.8&240.7&232.3&225.7&220.5\\
3.162&349.6&327.1&307.3&289.9&274.6&261.2&250.2&241.4&234.6&229.7\\
3.981&370.6&344.4&321.7&302.1&285.1&270.5&258.7&249.5&242.7&238.0\\
5.012&390.6&360.4&334.7&312.9&294.5&279.2&266.9&257.4&250.4&245.8\\
6.310&409.4&374.7&346.0&322.3&302.9&287.2&274.7&265.0&257.9&253.1\\
7.943&427.2&387.7&355.7&330.2&309.9&294.2&281.6&271.8&264.3&259.2\\
10.00&444.5&399.4&363.9&336.4&315.4&299.8&287.2&277.1&269.2&263.5\\
12.59&460.8&409.7&370.5&340.9&319.2&303.6&291.0&280.6&272.3&265.8\\
15.85&475.6&418.2&375.2&343.5&321.0&305.6&293.0&282.3&273.3&266.0\\
19.95&488.7&424.9&378.0&344.3&321.0&305.7&293.0&281.9&272.3&264.1\\
25.12&500.0&429.6&378.9&343.1&319.1&303.9&291.1&279.6&269.3&260.2\\
31.62&509.6&433.1&379.0&341.4&316.7&301.5&288.5&276.6&265.7&255.7\\
39.81&517.6&436.4&379.5&340.4&315.0&299.7&286.5&274.2&262.7&252.1\\
50.12&523.9&439.4&380.5&340.2&314.2&298.4&284.9&272.3&260.4&249.3\\
63.10&528.4&442.2&382.0&340.8&314.1&297.8&283.9&270.8&258.7&247.4\\
79.43&531.2&444.6&383.9&342.1&314.7&297.8&283.3&270.0&257.7&246.3\\
100.0&532.1&446.7&386.3&344.2&316.2&298.4&283.2&269.6&257.2&246.1\\
125.9&531.7&448.0&388.1&345.9&317.2&298.4&282.6&268.6&256.3&245.3\\
158.5&530.4&448.0&388.5&346.1&316.6&296.9&280.4&266.0&253.6&242.9\\
199.5&528.2&446.8&387.4&344.6&314.5&293.7&276.6&261.9&249.3&238.7\\
251.2&525.2&444.3&384.9&341.6&310.8&289.0&271.2&256.2&243.5&233.0\\
316.2&521.3&440.6&380.9&337.1&305.5&282.9&264.5&249.0&236.2&225.8\\
398.1&516.4&435.1&374.9&330.6&298.4&275.2&256.3&240.7&227.8&217.4\\
501.2&510.3&427.2&366.1&321.4&289.0&265.7&246.8&231.3&218.6&208.3\\
631.0&503.0&417.3&355.0&309.9&277.6&254.6&236.1&220.9&208.5&198.5\\
794.3&494.6&405.4&341.7&296.4&264.5&242.3&224.5&209.8&197.8&188.2\\
1000.&485.3&391.7&326.5&281.1&250.0&228.9&212.0&198.1&186.7&177.4\\
1259.&474.4&375.7&308.8&263.5&233.4&213.7&198.1&185.1&174.3&165.3\\
1585.&461.6&356.9&288.2&243.2&214.3&196.5&182.2&170.2&160.0&151.5\\
1995.&447.1&335.8&265.4&220.9&193.7&177.8&165.0&154.1&144.6&136.5\\
2512.&431.1&313.0&241.2&197.6&172.2&158.3&147.1&137.3&128.6&120.9\\
%3162.&413.8&289.0&216.2&173.8&150.5&138.7&129.2&120.4&(112.5)&(105.2)\\
3162.&406.0&296.5&228.2&185.2&158.4&142.0&128.8&(117.5)&(107.8)&(99.44)\\
$10^4$&(301.2)&(188.1)&(127.1)&(93.0)&(73.6)&( 62.5)&(53.6)&(46.1)&(39.6)&(34.1)\\
$3.16\cdot 10^4$&(201.2)&(99.5)&(55.45)&(34.8)&(24.7)&(19.4)&(15.4)&(12.1)&(9.5)&(7.4)\\
\end{tabular}
\end{table}

\begin{table}
\caption{$\bar \nu_\mu-$flux $\times E_\nu^3$ 
(${\rm m^{-2} sec^{-1} sr^{-1} GeV^2}$) 
calculated with the hybrid method.
The value above $1\times 10^4$ is the smooth extension
with a power fit.
}
\label{s-numu-bar3}
\begin{tabular}{c c c c c c c c c c c}
%s-numu-bar.d3
 $E_\nu({\rm GeV})\backslash \cos\theta$&$ .0-.1$&$ .1-.2$&$ .2-.3$&$ .3-.4$&$ .4-.5$&$ .5-.6$&$ .6-.7$&$ .7-.8$&$ .8-.9$&$ .9-1.$\\
\hline
1.000&217.1&210.4&203.3&196.0&188.3&180.6&173.5&167.0&161.1&155.7\\
1.259&246.6&237.4&228.0&218.4&208.7&199.1&190.4&182.6&175.7&169.6\\
1.585&274.4&262.2&249.9&237.7&225.6&213.8&203.4&194.5&186.9&180.5\\
1.995&299.6&283.5&267.9&252.8&238.2&224.3&212.4&202.6&194.6&188.2\\
2.512&323.6&302.6&283.2&265.0&248.2&232.7&219.7&209.2&200.9&194.6\\
3.162&348.0&321.3&297.5&276.5&257.7&241.1&227.5&216.4&207.7&201.1\\
3.981&371.7&338.6&310.5&286.6&266.3&249.1&235.0&223.6&214.5&207.5\\
5.012&393.4&353.9&321.5&295.0&273.4&255.9&241.6&229.9&220.5&213.2\\
6.310&412.6&366.6&330.2&301.4&278.9&261.5&247.1&235.3&225.6&218.0\\
7.943&429.2&377.0&336.7&305.9&282.7&265.4&251.1&239.2&229.4&221.4\\
10.00&443.5&385.1&341.2&308.4&284.5&267.3&253.2&241.2&231.2&223.0\\
12.59&455.6&391.0&343.5&308.9&284.3&267.3&253.3&241.3&231.2&222.7\\
15.85&465.5&394.8&343.7&307.4&282.3&265.6&251.7&239.7&229.4&220.5\\
19.95&473.2&396.2&341.8&303.9&278.4&262.1&248.5&236.5&225.9&216.6\\
25.12&478.4&395.3&337.8&298.5&272.8&257.0&243.6&231.6&220.8&211.1\\
31.62&481.3&393.0&332.8&292.5&266.7&251.3&238.2&226.3&215.3&205.2\\
39.81&482.1&389.9&328.0&286.9&261.2&246.1&233.3&221.4&210.3&200.0\\
50.12&480.9&386.3&323.3&281.9&256.2&241.5&228.9&217.1&206.0&195.6\\
63.10&477.6&382.0&318.7&277.3&251.8&237.3&224.9&213.2&202.2&191.8\\
79.43&472.2&377.1&314.2&273.2&247.9&233.5&221.3&209.8&198.9&188.7\\
100.0&464.8&371.6&309.8&269.4&244.4&230.2&218.1&206.8&196.2&186.2\\
125.9&456.0&365.2&304.8&265.2&240.6&226.5&214.5&203.4&193.0&183.4\\
158.5&446.2&357.5&298.4&259.6&235.4&221.5&209.7&198.8&188.7&179.3\\
199.5&435.4&348.6&290.8&252.8&229.1&215.4&203.8&193.1&183.2&174.0\\
251.2&423.8&338.6&282.0&244.8&221.6&208.2&196.9&186.4&176.6&167.6\\
316.2&411.5&327.7&272.1&235.7&213.2&200.1&189.1&178.8&169.2&160.3\\
398.1&398.7&316.4&262.0&226.4&204.3&191.5&180.6&170.4&160.9&152.0\\
501.2&385.6&305.5&252.4&217.5&195.7&182.7&171.6&161.3&151.6&142.6\\
631.0&372.3&294.9&243.2&209.1&187.2&173.8&162.3&151.6&141.7&132.5\\
794.3&358.9&284.6&234.6&201.0&179.0&164.9&152.8&141.6&131.3&121.9\\
1000.&345.5&274.7&226.4&193.2&170.9&156.0&143.1&131.3&120.7&110.9\\
1259.&331.6&264.6&217.9&185.2&162.4&146.5&132.7&120.4&109.3&(99.4)\\
1585.&317.1&253.8&208.7&176.3&153.0&136.0&121.4&108.5&(97.2)&(87.1)\\
1995.&302.1&242.4&198.8&166.6&142.8&124.8&109.5&(96.2)&(84.7)&(74.7)\\
2512.&286.7&230.6&188.2&156.3&132.0&113.2&(97.4)&(83.9)&(72.4)&(62.6)\\
%3162.&271.0&218.3&177.2&145.5&121.0&101.5&(85.4)&(72.0)&(60.7)&(51.3)\\
3162.&278.89&210.88&165.26&134.27&113.09&(98.3)&(85.8)&(74.9)&(65.5)&(57.2)\\

$10^4$&(206.9)&(133.8)&(92.1)&(67.4)&(52.6)&(43.2)&(35.7)&(29.4)&(24.1)&(19.6)\\
$3.16\cdot 10^4$&(138.2)&(70.8)&(40.2)&(25.3)&(17.6)&(13.4)&(10.2)&(7.7)&(5.8)&(4.3)\\
\end{tabular}
\end{table}

\begin{table}
\caption{$\nu_e-$flux $\times E_\nu^3$ 
 (${\rm m^{-2} sec^{-1} sr^{-1} GeV^2}$) 
calculated with the hybrid method.}
\label{s-nue3}
\begin{tabular}{c c c c c c c c c c c}
%s-nue.d3
$E_\nu({\rm GeV})\backslash \cos\theta$&$ .0-.1$&$ .1-.2$&$ .2-.3$&$ 
.3-.4$&$ .4-.5$&$ .5-.6$&$ .6-.7$&$ .7-.8$&$ .8-.9$&$ .9-1.$\\
\hline
1.000&131.2&124.3&116.7&109.0&101.3& 94.0&87.2&81.0&75.6&70.9\\
1.259&150.9&140.8&130.5&120.4&110.7&101.7&93.6&86.4&80.2&74.9\\
1.585&166.3&154.8&142.2&129.5&117.2&105.9&95.9&87.4&80.4&75.0\\
1.995&179.2&165.4&150.1&134.7&120.1&106.9&95.5&86.0&78.6&73.1\\
2.512&192.5&172.0&152.6&134.9&119.1&105.5&94.0&84.6&77.1&71.3\\
3.162&205.4&176.7&152.7&132.7&116.2&102.5&91.2&81.9&74.4&68.4\\
3.981&216.8&181.7&153.6&131.1&113.0&98.4&86.6&77.1&69.4&63.2\\
5.012&226.2&186.9&155.3&129.9&109.7&93.5&80.7&70.5&62.5&56.3\\
6.310&233.6&190.4&155.5&127.5&105.3&87.8&74.2&63.7&55.6&49.7\\
7.943&239.0&190.2&151.9&122.1&99.1&81.4&67.9&57.6&50.0&44.3\\
10.00&242.3&186.2&144.9&114.2&91.5&74.5&61.8&52.3&45.3&40.1\\
12.59&243.5&178.8&134.8&104.3&82.8&67.3&55.9&47.6&41.4&36.8\\
15.85&242.5&168.3&122.5&93.0&73.4&59.9&50.4&43.4&38.2&34.1\\
19.95&239.1&156.4&109.8&82.0&64.6&53.1&45.3&39.7&35.4&31.8\\
25.12&233.3&144.7&98.4&72.4&56.9&47.2&40.8&36.2&32.7&29.5\\
31.62&225.2&133.1&88.0&63.9&50.2&42.0&36.7&33.0&30.1&27.3\\
39.81&215.2&121.9&78.6&56.5&44.4&37.4&33.1&30.1&27.6&25.1\\
50.12&203.6&111.2&70.2&50.1&39.5&33.5&29.9&27.5&25.3&22.9\\
63.10&190.7&100.9&62.7&44.5&35.1&30.0&27.1&25.0&23.1&20.7\\
79.43&176.8&91.3&56.0&39.6&31.4&27.0&24.5&22.8&21.0&18.7\\
100.0&162.3&82.2&50.0&35.3&28.1&24.3&22.2&20.7&19.0&16.7\\
125.9&147.4&73.6&44.6&31.5&25.2&22.0&20.2&18.8&17.2&14.9\\
158.5&132.6&65.7&39.7&28.2&22.7&19.9&18.3&17.1&15.5&13.1\\
199.5&118.1&58.4&35.4&25.3&20.5&18.1&16.7&15.5&13.9&(11.5)\\
251.2&104.1&51.7&31.6&22.8&18.6&16.5&15.2&14.0&12.4&(10.0)\\
316.2&90.9&45.5&28.1&20.5&16.9&15.1&13.9&12.7&(11.0)&(8.7)\\
398.1&78.5&39.9&25.0&18.5&15.4&13.8&12.7&(11.5)&(9.8)&(7.5)\\
501.2&67.1&34.9&22.3&16.7&14.1&(12.7)&(11.6)&(10.4)&(8.7)&(6.4)\\
631.0&56.8&30.3&19.8&15.1&12.9&(11.6)&(10.6)&(9.4)&(7.6)&(5.4)\\
794.3&47.6&26.2&17.6&13.7&(11.8)&(10.7)&(9.8)&(8.5)&(6.7)&(4.6)\\
1000.&39.5&22.6&15.7&12.5&(10.9)&(9.9)&(9.0)&(7.7)&(5.9)&(3.8)\\
\end{tabular}
\end{table}

\begin{table}
\caption{$\bar\nu_e-$flux $\times E_\nu^3$ 
(${\rm m^{-2} sec^{-1} sr^{-1} GeV^2}$) calculated with the hybrid method.}
\label{s-nue-bar3}
\begin{tabular}{c c c c c c c c c c c}
%s-nue-bar.d3
 $E_\nu({\rm GeV})\backslash \cos\theta$&$ .0-.1$&$ .1-.2$&$ .2-.3$&$ .3-.4$&$ .4-.5$&$ .5-.6$&$ .6-.7$&$ .7-.8$&$ .8-.9$&$ .9-1.$\\
\hline
1.000&109.2&103.1& 96.5& 89.8&83.2&77.0&71.1&65.9&61.3&57.3\\
1.259&126.1&117.3&108.3& 99.6&91.4&83.7&76.8&70.7&65.4&61.0\\
1.585&139.2&129.2&118.4&107.5&97.1&87.5&79.1&71.9&66.1&61.5\\
1.995&149.9&138.1&125.1&112.0&99.7&88.7&79.1&71.2&65.0&60.5\\
2.512&160.7&143.4&127.1&112.2&99.1&87.8&78.3&70.5&64.3&59.6\\
3.162&170.7&146.7&126.8&110.3&96.7&85.4&76.2&68.7&62.6&57.8\\
3.981&179.2&150.2&127.2&108.8&94.1&82.2&72.7&65.0&58.8&53.9\\
5.012&186.0&154.0&128.3&107.8&91.3&78.2&67.9&59.7&53.3&48.3\\
6.310&191.2&156.4&128.2&105.7&87.7&73.6&62.5&54.0&47.5&42.7\\
7.943&194.8&155.8&125.1&101.2&82.6&68.3&57.3&48.9&42.7&38.2\\
10.00&196.8&152.3&119.2& 94.6&76.2&62.5&52.1&44.4&38.6&34.4\\
12.59&197.1&146.0&111.0& 86.5&69.0&56.4&47.2&40.3&35.2&31.3\\
15.85&195.8&137.3&100.8& 77.1&61.3&50.3&42.4&36.6&32.2&28.8\\
19.95&192.6&127.6& 90.4& 68.1&53.9&44.5&38.0&33.3&29.6&26.5\\
25.12&187.6&118.0& 81.1& 60.1&47.5&39.5&34.1&30.2&27.0&24.2\\
31.62&180.9&108.6& 72.6& 53.2&42.0&35.1&30.6&27.3&24.6&21.9\\
39.81&172.6& 99.5& 65.0& 47.1&37.1&31.2&27.4&24.7&22.3&19.7\\
50.12&163.0& 90.6& 58.0& 41.7&32.9&27.8&24.6&22.2&20.1&17.7\\
63.10&152.3& 82.1& 51.6& 36.9&29.1&24.7&22.0&20.0&18.1&15.7\\
79.43&140.8& 74.0& 45.9& 32.6&25.8&22.1&19.8&18.0&16.2&13.9\\
100.0&128.8& 66.3& 40.7& 28.9&22.9&19.7&17.7&16.1&14.4&12.3\\
125.9&116.5& 59.0& 36.0& 25.5&20.3&17.6&15.9&14.5&12.8&(10.7)\\
158.5&104.4& 52.3& 31.8& 22.6&18.1&15.7&14.2&12.9&11.4&(9.4)\\
199.5& 92.5& 46.1& 28.0& 20.0&16.1&14.0&12.7&11.5&(10.1)&(8.1)\\
251.2& 81.1& 40.3& 24.6& 17.7&14.3&12.5&11.4&(10.3)&(8.9)&(7.0)\\
316.2& 70.3& 35.1& 21.6& 15.6&12.7&11.2&(10.2)&(9.1)&(7.8)&(6.0)\\
398.1& 60.3& 30.4& 18.9& 13.8&11.3&(10.0)&(9.1)&(8.1)&(6.8)&(5.1)\\
501.2& 51.2& 26.2& 16.5& 12.2&(10.1)&(9.0)&(8.1)&(7.2)&(5.9)&(4.3)\\
631.0& 43.0& 22.4& 14.4&(10.8)&(9.0)&(8.0)&(7.3)&(6.4)&(5.1)&(3.6)\\
794.3& 35.8& 19.1& 12.5& (9.5)&(8.0)&(7.2)&(6.5)&(5.6)&(4.4)&(3.0\\
1000.& 29.4& 16.2&(10.8)&(8.4)&(7.2)&(6.4)&(5.8)&(4.9)&(3.8)&(2.5)\\
\end{tabular}
\end{table}
\vspace{1pc}

\end{document}